\documentclass[12pt]{article}
\usepackage{graphicx}
\def\beq{\begin{equation}}
\def\eeq{\end{equation}}
\def\bea{\begin{eqnarray}}
\def\eea{\end{eqnarray}}
\def\nn{\nonumber}

\def\roughly#1{\mathrel{\raise.3ex\hbox
{$#1$\kern-.75em\lower1ex\hbox{$\sim$}}}}

\def\sla#1{\raise.15ex\hbox{$/$}\kern-.57em #1}
\def\bra#1{\left\langle #1\right|}
\def\ket#1{\left| #1\right\rangle}
\def\ks{K_S}
\def\kbar{{\bar K}^0}
\def\bd{B_d^0}
\def\bdbar{{\bar B}_d^0}
\def\pewc{P_{EW}^C}
\def\pew{P_{EW}}

\def\btod{{\bar b} \to {\bar d}}
\def\btos{{\bar b} \to {\bar s}}

\begin{document}

\begin{flushright}
UdeM-GPP-TH-10-194 \\
\end{flushright}

\begin{center}
\bigskip
{\Large \bf \boldmath Diagrammatic Analysis of Charmless \\ Three-Body $B$ Decays} \\
\bigskip
\bigskip
{\large 
Nicolas Rey-Le Lorier $^{a,}$\footnote{nicolas.rey-le.lorier@umontreal.ca},
Maxime Imbeault $^{b,}$\footnote{imbeault.maxime@gmail.com},
and David London $^{a,}$\footnote{london@lps.umontreal.ca}
}
\end{center}

\begin{flushleft}
~~~~~~~~~~~$a$: {\it Physique des Particules, Universit\'e
de Montr\'eal,}\\
~~~~~~~~~~~~~~~{\it C.P. 6128, succ. centre-ville, Montr\'eal, QC,
Canada H3C 3J7}\\
~~~~~~~~~~~$b$: {\it D\'epartement de physique, C\'egep de Baie-Comeau,}\\
~~~~~~~~~~~~~~~{\it 537 boulevard Blanche, Baie-Comeau, QC, Canada G5C 2B2}\\
\end{flushleft}

\begin{center}
\bigskip (\today)
\vskip0.5cm {\Large Abstract\\} \vskip3truemm
\parbox[t]{\textwidth}{We express the amplitudes for charmless
  three-body $B$ decays in terms of diagrams.  In addition, we show
  how to use Dalitz-plot analyses to obtain decay amplitudes which are
  symmetric or antisymmetric under the exchange of two of the
  final-state particles.  When annihilation-type diagrams are
  neglected, as in two-body decays, many of the exact, purely
  isospin-based results are modified, leading to new tests of the
  standard model (SM). Some of the tests can be performed now, and we
  find that present data agree with the predictions of the SM.
  Furthermore, contrary to what was thought previously, it is possible
  to cleanly extract weak-phase information from three-body decays,
  and we discuss methods for $B \to K\pi\pi$, $KK{\bar K}$, $K{\bar
    K}\pi$ and $\pi\pi\pi$.  }
\end{center}

\thispagestyle{empty}
\newpage
\setcounter{page}{1}
\baselineskip=14pt

\section{Introduction}

The $B$-factories BaBar and Belle ran for over ten years, and made an
enormous number of measurements of observables in $B$ decays. For the
most part, these decays were of the form $B \to M_1 M_2$ ($M_i$ is a
meson), as these are most accessible experimentally. Nevertheless,
there have still been some probes of three-body $B \to M_1 M_2 M_3$
decays. To be specific, experiments have obtained Dalitz plots for
many of the decay modes in $B \to K\pi\pi$, $KK{\bar K}$, $K{\bar
  K}\pi$, $\pi\pi\pi$, and made measurements of (or obtained upper
limits on) the branching ratios and indirect (mixing-induced) CP
asymmetries of a number of these decays \cite{hfag}.

Things are similar on the theory side. The vast majority of
theoretical analyses involve two-body $B$ decays. This is in part due
to the relative angular momentum of the final-state particles. For
example, consider $\bd \to \pi^+\pi^-$. Because there are two
particles in the final state, it has a fixed value of $l$ (in this
case $l=0$), and so $\pi^+\pi^-$ is a CP eigenstate. On the other
hand, in the decay $\bd \to \ks \pi^+\pi^-$, the $\pi^+\pi^-$ can have
even or odd relative angular momentum, so that $\ks \pi^+\pi^-$ is not
a CP eigenstate. This makes it much more difficult to find clean
predictions of the standard model (SM) to compare with experimental
measurements.  This is a general property of three-body decays.

Still, there have been some theoretical analyses of CP-conserving
observables in three-body $B \to K\pi\pi$, $KK{\bar K}$ decays
\cite{LNQS,GR2003,GR2005,Sonietal}. In general, these studies examined
the isospin decomposition of the decay amplitudes, and symmetry
relations among them. The analyses were carried out using isospin
amplitudes.

In this paper, we examine the amplitudes of the three-body charmless
decays $B \to K\pi\pi$, $KK{\bar K}$, $K{\bar K}\pi$, $\pi\pi\pi$
using diagrams. In addition, using Dalitz-plot analyses of such
decays, we show how to separate the amplitudes into pieces which are
symmetric or antisymmetric under the exchange of two of the
final-state particles. This is useful for any decay which contains
particles which are identical under isospin.  Now, as has been shown
in Ref.~\cite{GHLR}, the amplitudes for two-body $B$ decays can be
expressed in terms of 9 diagrams.  However, 3 of these -- the
annihilation-type diagrams -- are expected to be quite a bit smaller
than the others, and can be neglected, to a good approximation. This
same procedure can be applied to three-body decays.

The point of this is as follows. When one neglects annihilation-type
diagrams, new features appear. A given set of three-body decays
(e.g.\ $B \to K\pi\pi$) contains a number of different transitions
(e.g.\ $B^+ \to K^+\pi^+\pi^-$, $\bd \to K^+\pi^0\pi^-$, etc.). There
are exact relations among the symmetric or antisymmetric amplitudes
for these specific decays. However, when one neglects certain
diagrams, these relations can be modified, and this can lead to new
effects. For example, some linear combinations of the isospin
amplitudes vanish for certain decays. Also, there are additional tests
of the SM. In some cases, it is even possible to obtain clean
information about the CP-violating phases.

In Sec.~2, we present the diagrams describing $B \to M_1 M_2 M_3$
processes. We review Dalitz-plot analyses of three-body decays in
Sec.~3, and show how to obtain amplitudes which are symmetric or
antisymmetric under the exchange of two of the final-state particles.
The decays $B \to K\pi\pi$, $B \to KK{\bar K}$, $B \to K{\bar K}\pi$
and $B \to \pi\pi\pi$ are discussed in Secs.~4, 5, 6 and 7,
respectively.  In all cases, we give the expressions for the decay
amplitudes in terms of diagrams, and examine the prospects for the
clean extraction of weak-phase information. Other subjects related to
the particular decays are also discussed: resonances and penguin
dominance in $B \to K\pi\pi$ (Sec.~4), penguin dominance and isospin
amplitudes in $B \to KK{\bar K}$ (Sec.~5), $T$ dominance in $B \to
K{\bar K}\pi$ (Sec.~6), and Dalitz plots in $B \to \pi\pi\pi$
(Sec.~7). We conclude in Sec.~8.

\section{Diagrams}

It has been shown in Ref.~\cite{GHLR} that the amplitudes for two-body
$B$ decays can be expressed in terms of 9 diagrams: the color-favored
and color-suppressed tree amplitudes $T$ and $C$, the gluonic-penguin
amplitudes $P_{tc}$ and $P_{uc}$, the color-favored and
color-suppressed electroweak-penguin (EWP) amplitudes $\pew$ and
$\pewc$, the annihilation amplitude $A$, the exchange amplitude $E$,
and the penguin-annihilation amplitude $PA$. These last three all
involve the interaction of the spectator quark, and are expected to be
much smaller than the other diagrams. It is standard to neglect them.
(Note that the neglect of such diagrams is justified experimentally --
no annihilation-type or exchange-type decays, such as $\bd \to
\phi\phi$, $B^+\to D_s \phi$, etc., have been observed \cite{hfag}.)

\begin{figure}
	\centering
		\includegraphics[height=3.98cm]{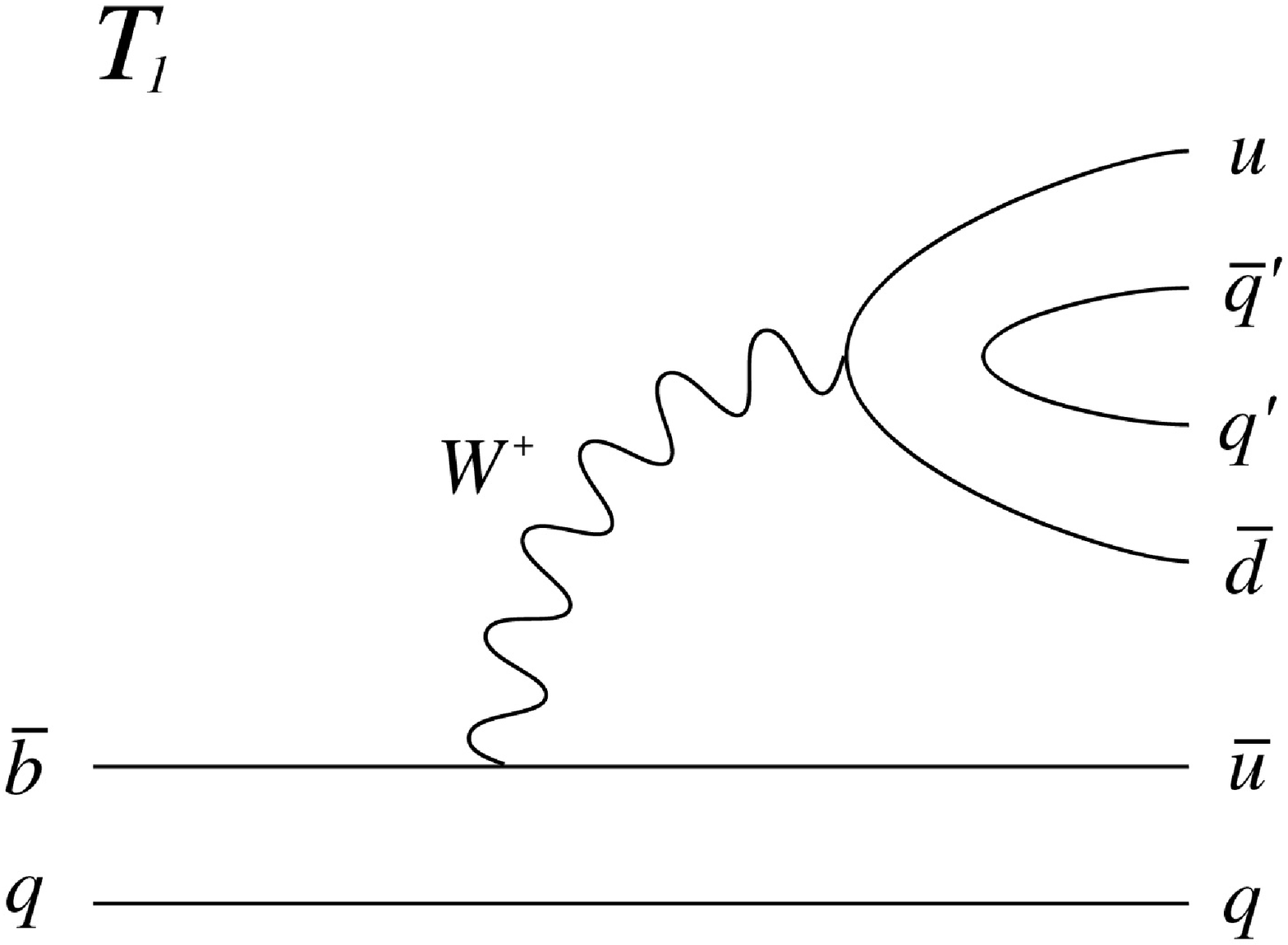}
		\includegraphics[height=3.98cm]{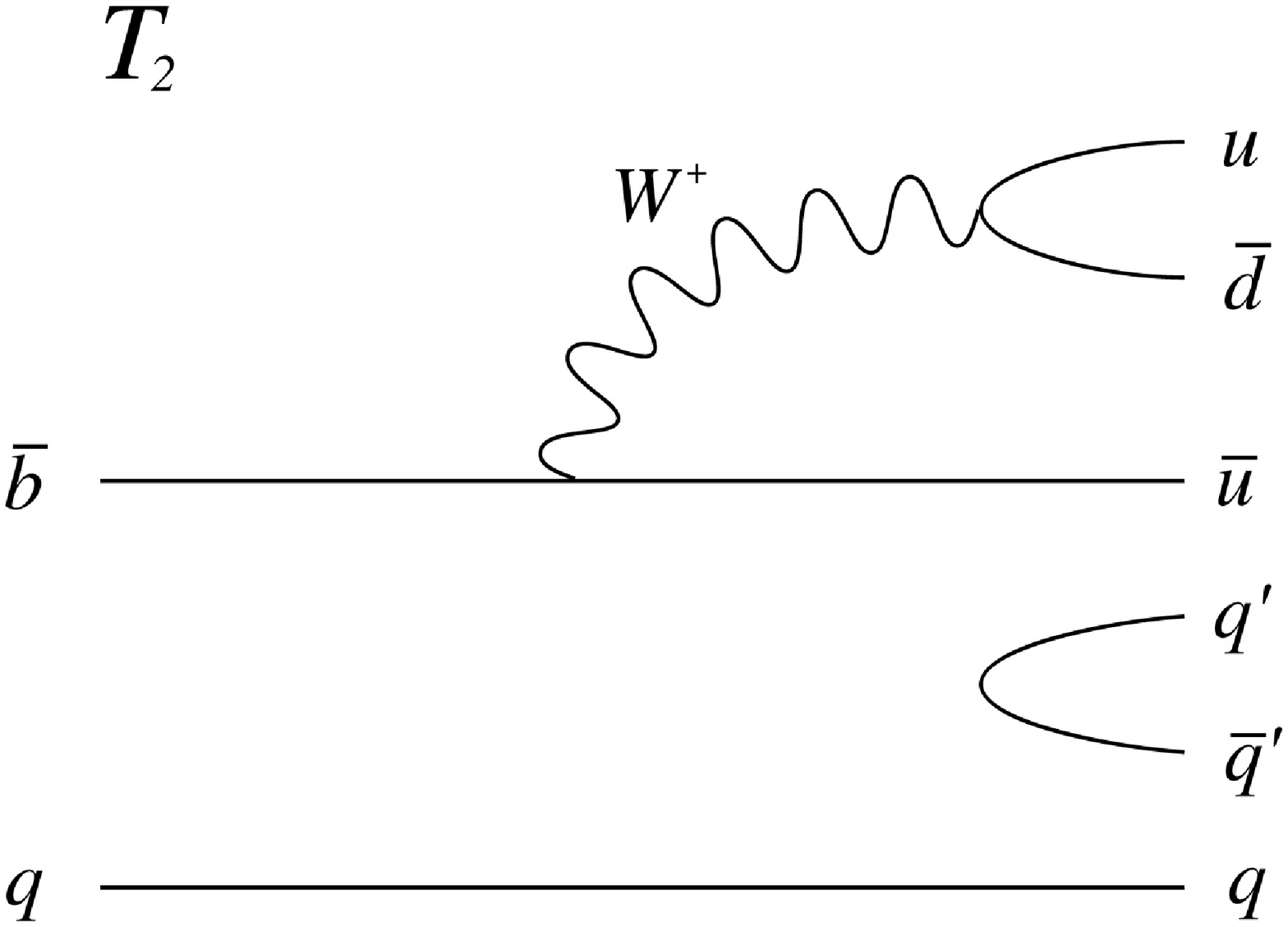}
	\centering
		\includegraphics[height=3.98cm]{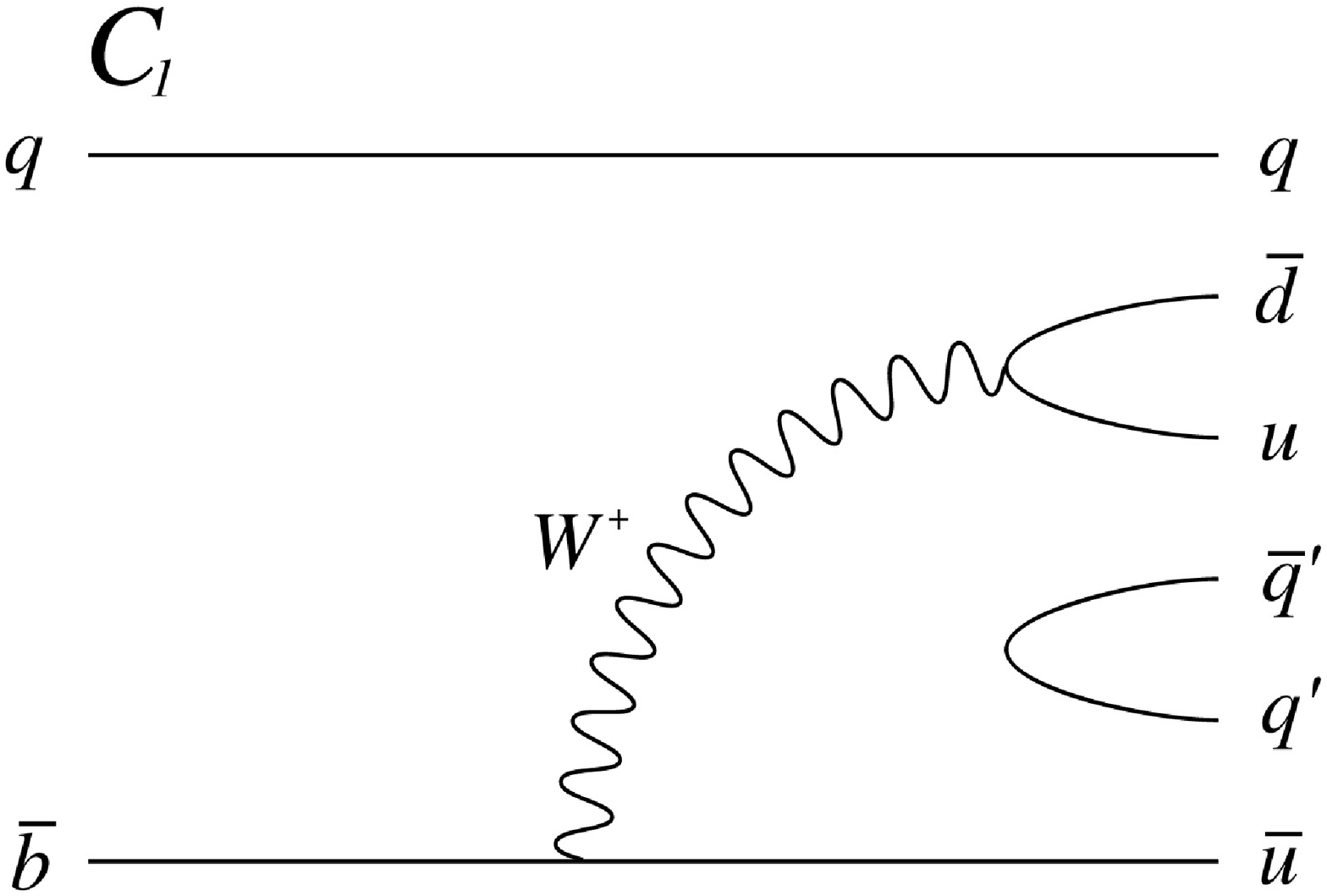}
		\includegraphics[height=3.98cm]{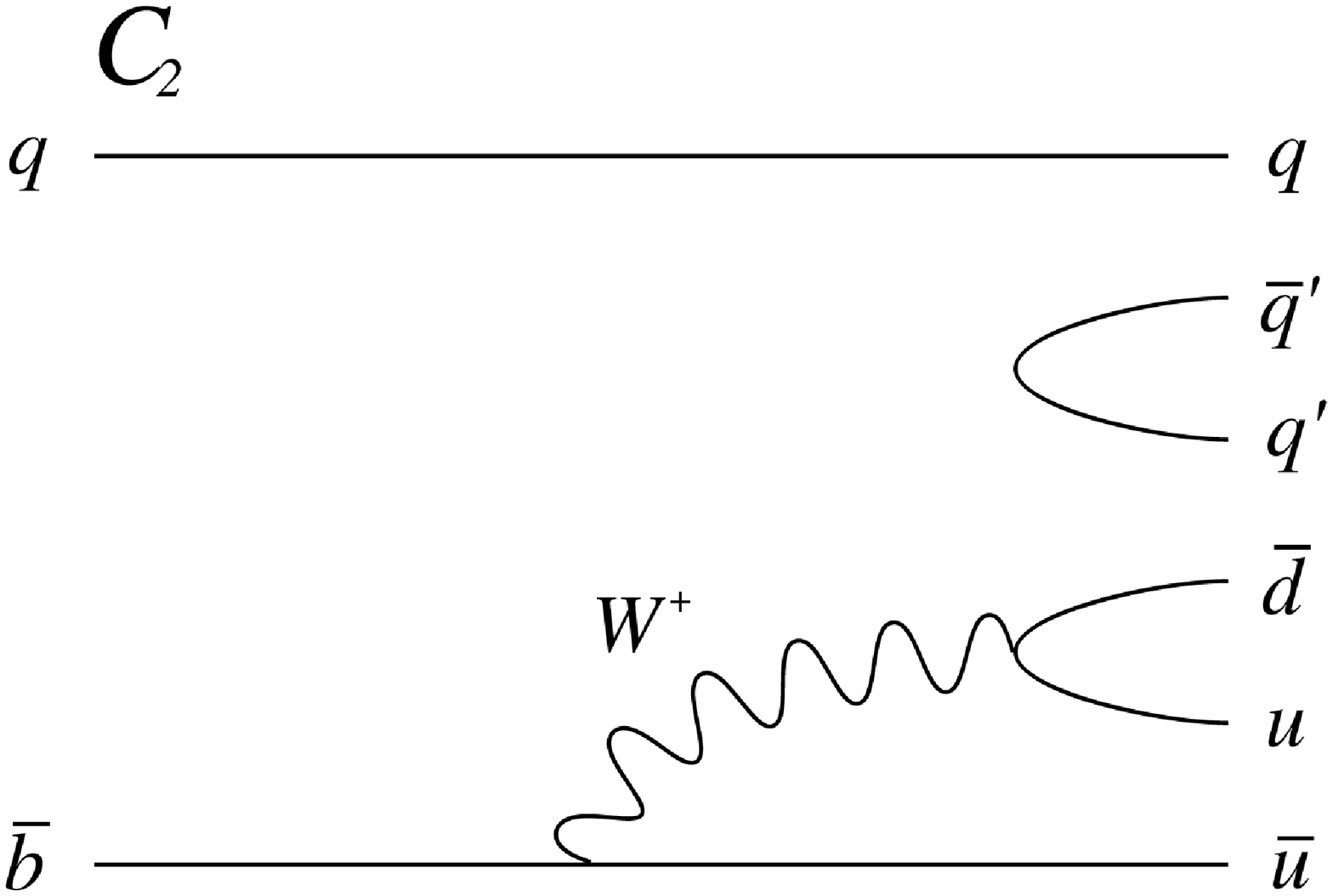}
	\centering
		\includegraphics[height=3.98cm]{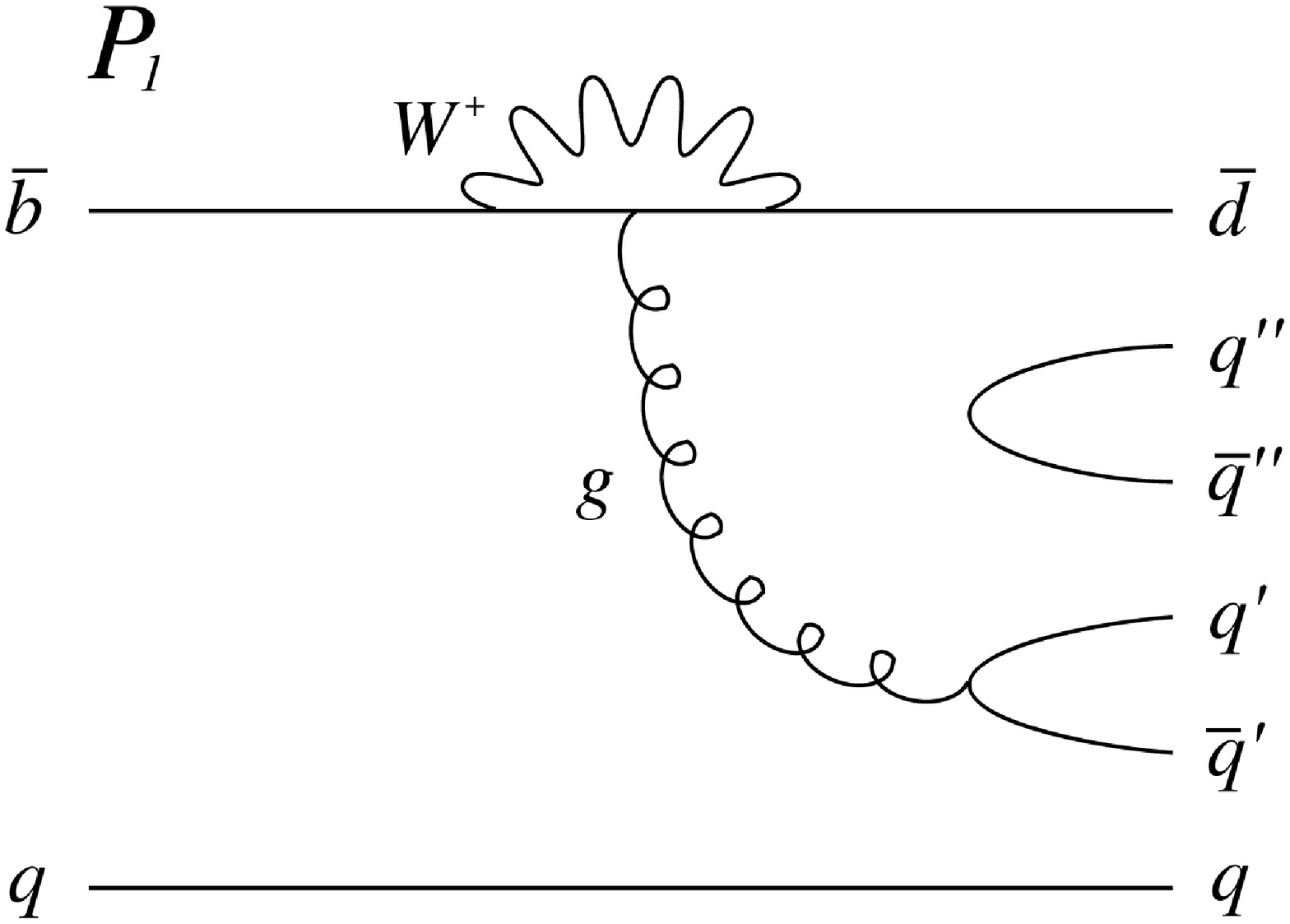}
		\includegraphics[height=3.98cm]{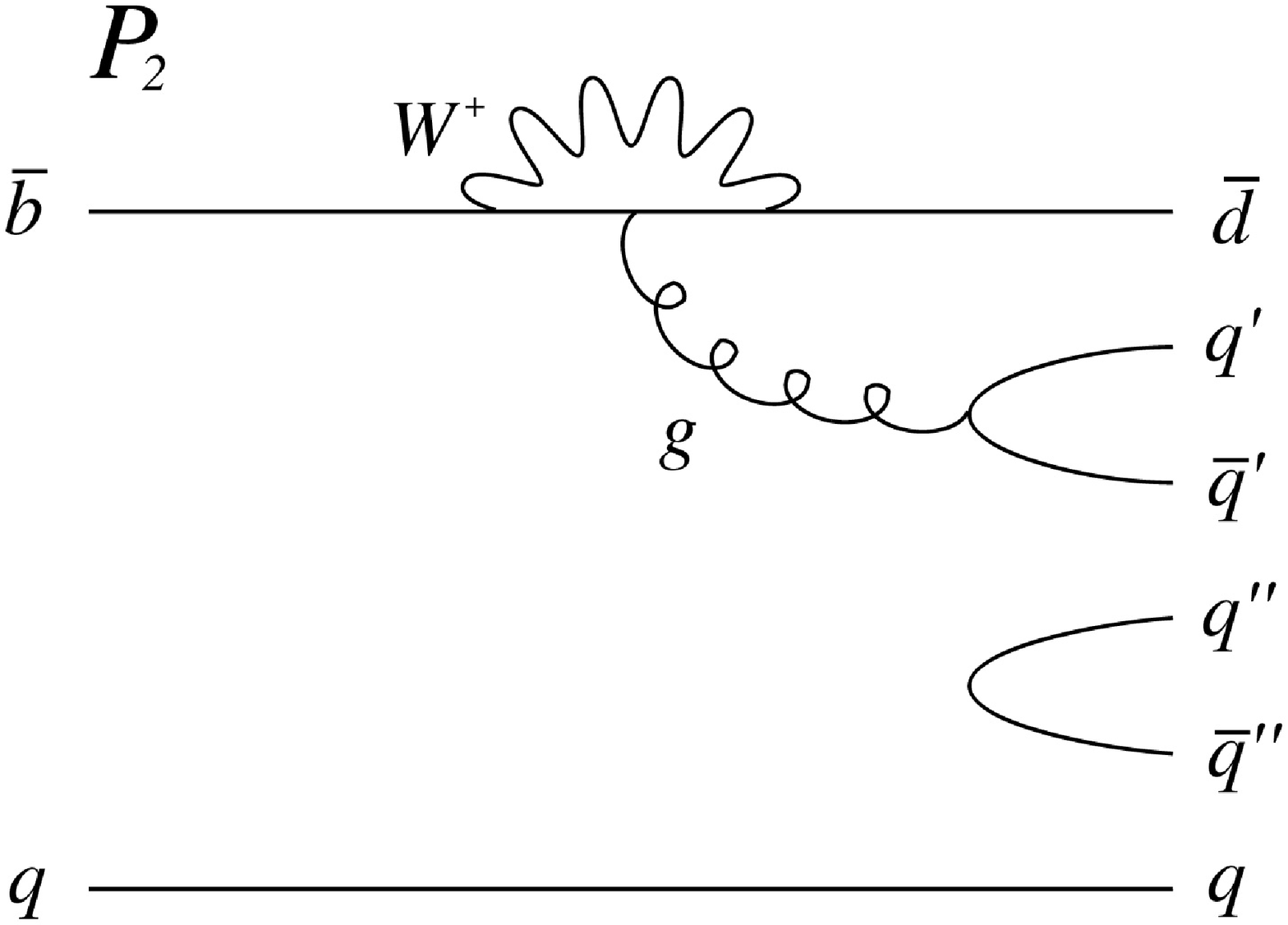}
	\centering
		\includegraphics[height=3.98cm]{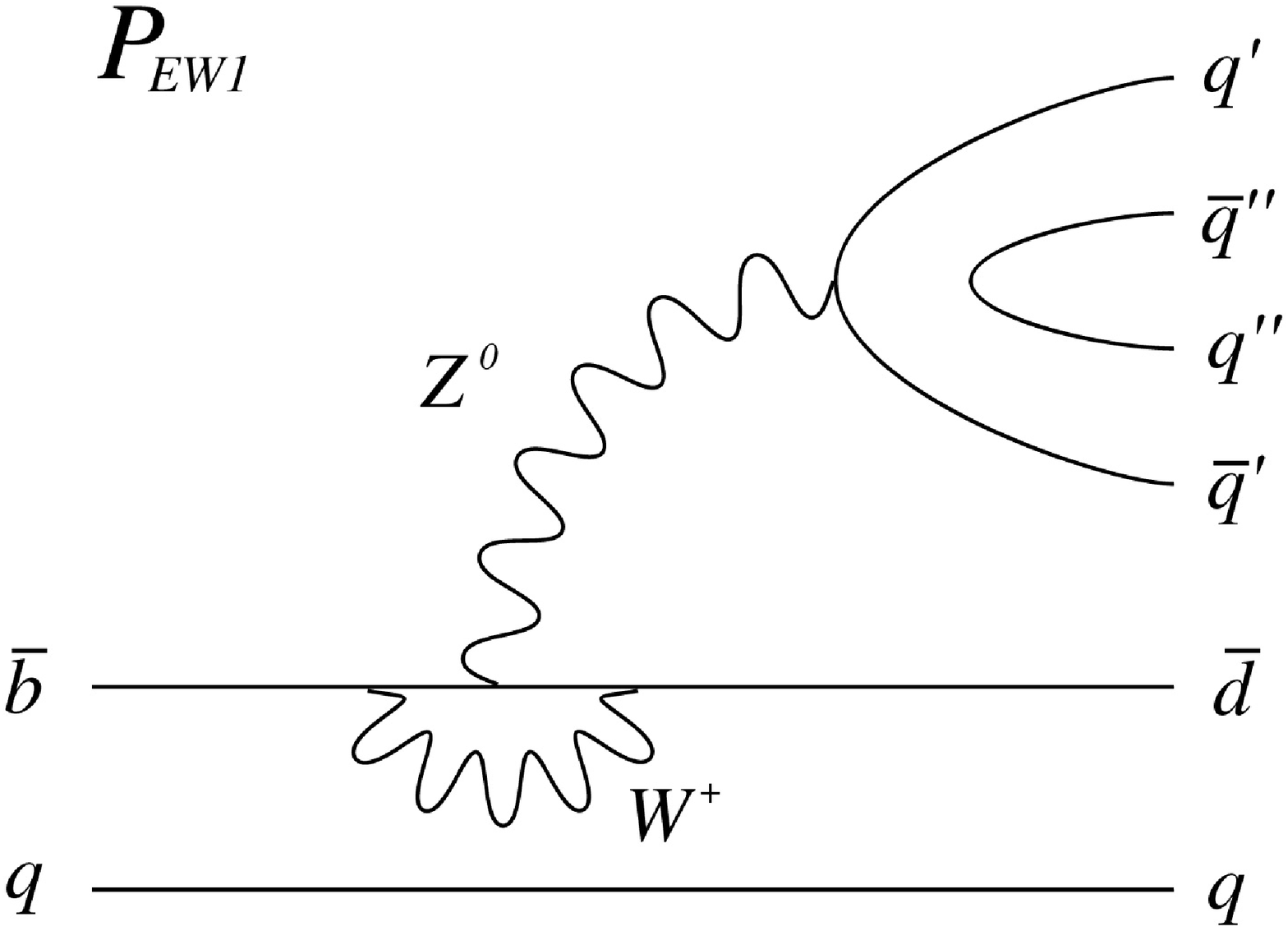}
		\includegraphics[height=3.98cm]{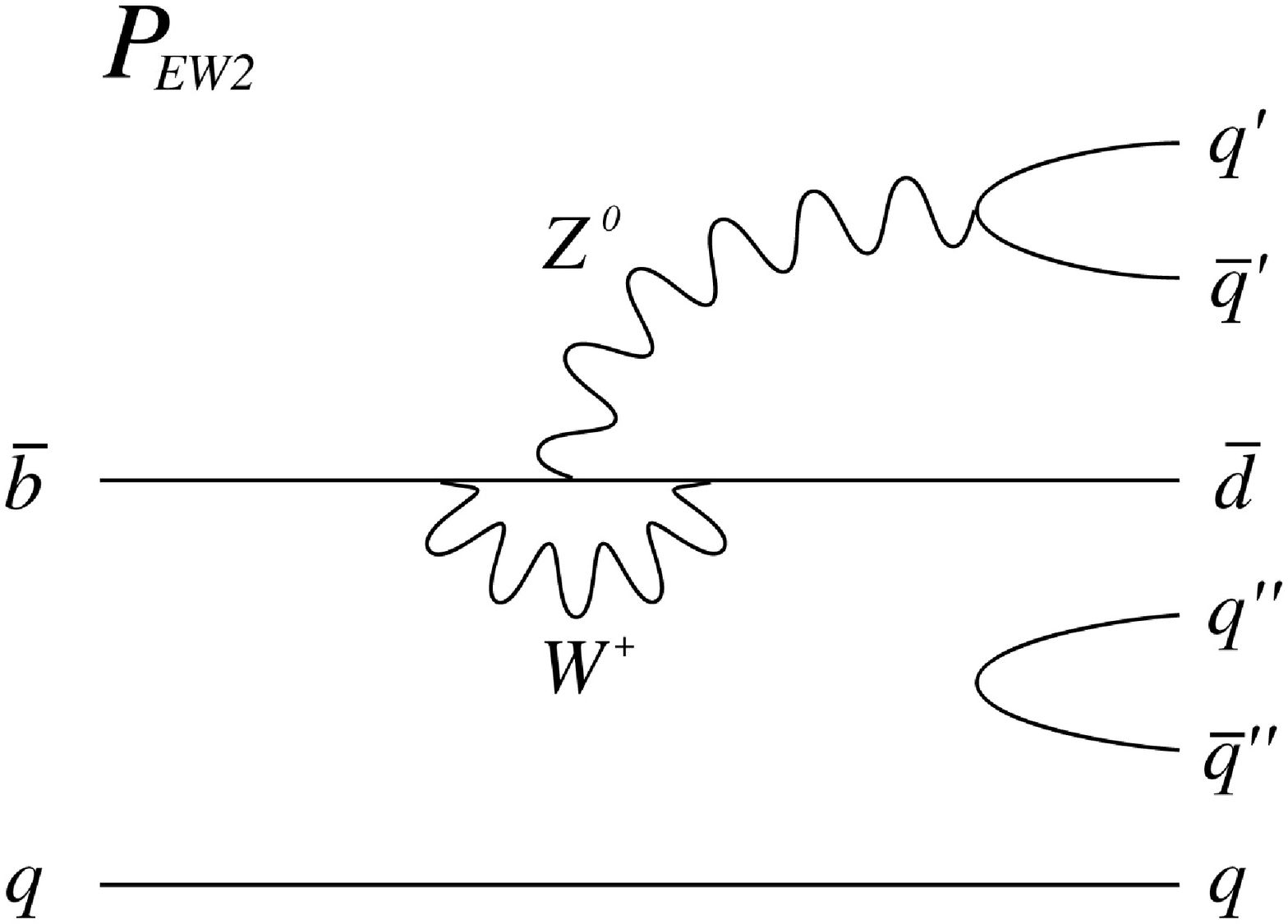}
	\centering
		\includegraphics[height=3.98cm]{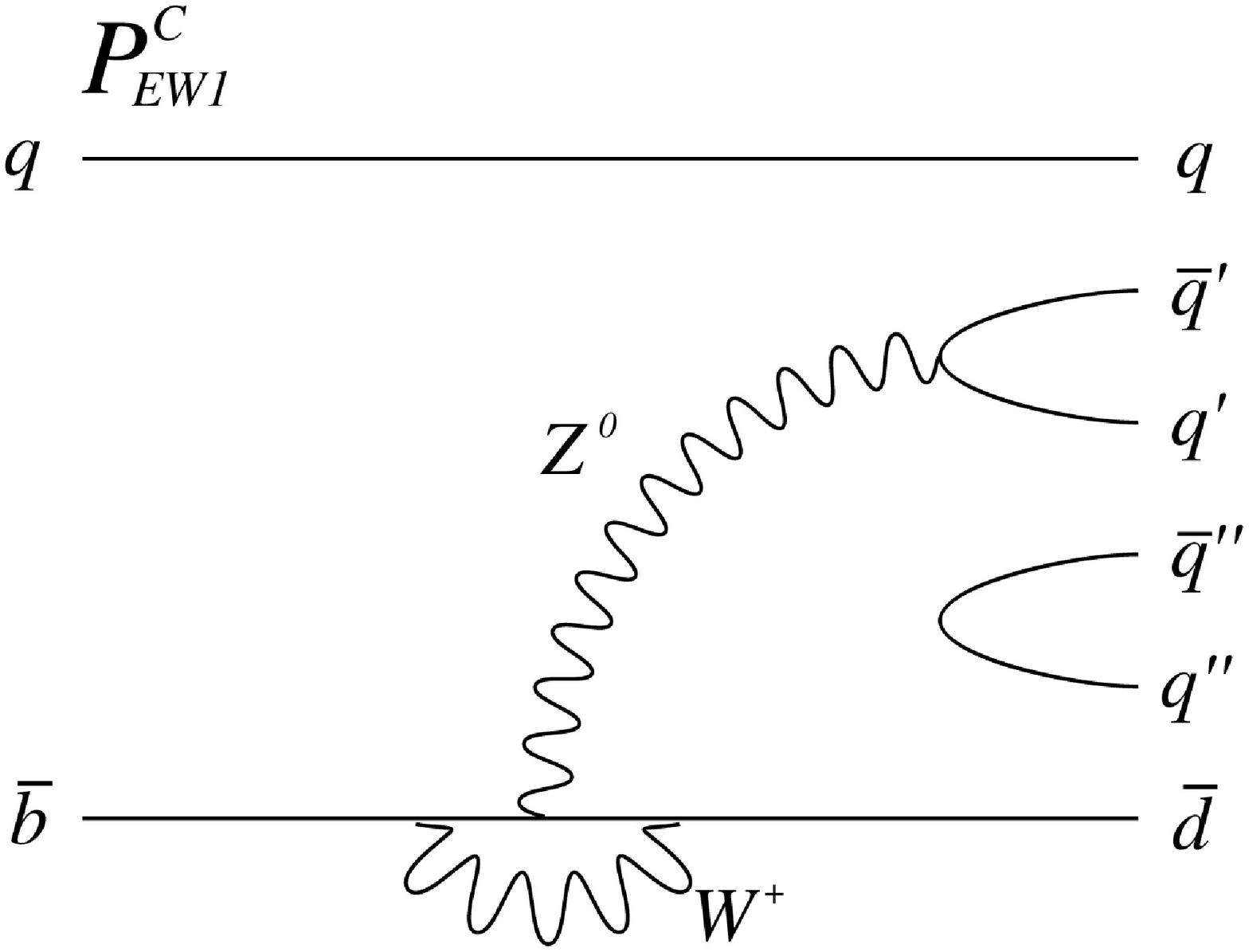}
		\includegraphics[height=3.98cm]{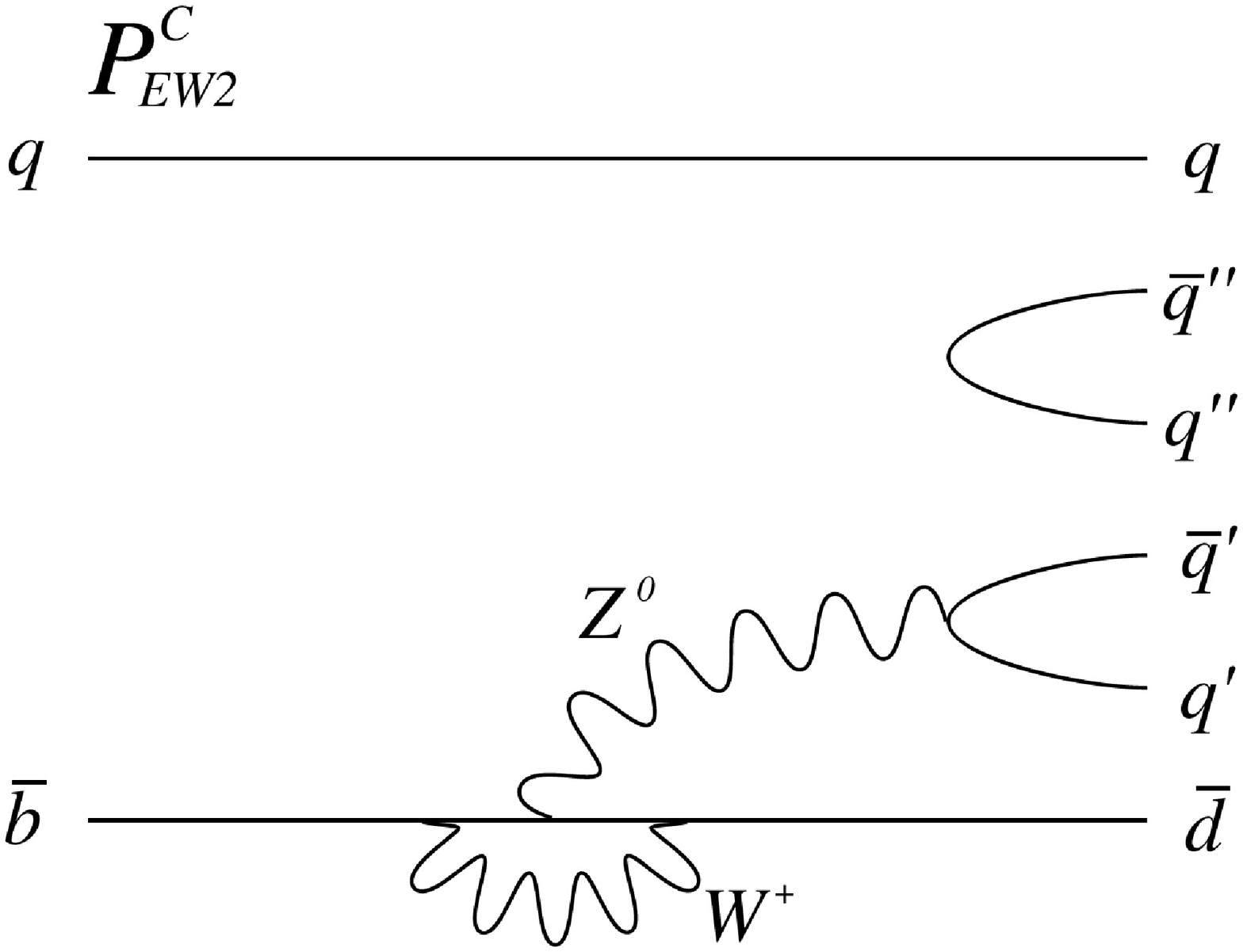}
\caption{Diagrams contributing to $B \to \pi\pi\pi$.}
\label{BPPPfig}
\end{figure}

For the three-body decays considered in this paper, we adopt a similar
procedure. That is, we neglect all annihilation-type diagrams, and
express all amplitudes in terms of tree, penguin, and EWP diagrams. We
assume isospin invariance, but not flavor SU(3) symmetry. (It is
straightforward to modify our analysis by imposing SU(3).) The
diagrams are shown in Fig.~\ref{BPPPfig}. A few words of
explanation. These diagrams are for the decay $B \to \pi\pi\pi$. There
are changes of notation for the other decays:
\begin{itemize}

\item For $\btod$ transitions ($B \to K{\bar K}\pi$, $\pi\pi\pi$), the
  diagrams are written without primes; for $\btos$ transitions ($B \to
  K\pi\pi$, $KK{\bar K}$), they are written with primes.

\item In all diagrams, it is necessary to ``pop'' a quark pair from
  the vacuum. It is assumed that this pair is $u{\bar u}$ or $d{\bar
    d}$ ($\equiv q {\bar q}$); if the popped pair is $s{\bar s}$, the
  diagram is written with an additional subscript ``$s$.'' Thus, for
  $B \to K{\bar K}\pi$, $KK{\bar K}$, in the penguin or EWP diagrams
  with a popped $q {\bar q}$ pair, the virtual particle decays to
  $s{\bar s}$; if the popped quark pair is $s{\bar s}$ (so the diagram
  is written with an additional subscript ``$s$''), the virtual
  particle decays to $q {\bar q}$.

\item The subscript ``1'' indicates that the popped quark pair is
  between two (non-spectator) final-state quarks; the subscript ``2''
  indicates that the popped quark pair is between two final-state
  quarks including the spectator.

\end{itemize}
In principle, one can also include the gluonic-penguin diagrams in
which the popped quark pair is between the pair of quarks produced by
the gluon. This corresponds to the case where the virtual spin-1 gluon
decays to two spin-0 mesons (with relative angular momentum $l=1$). In
order to account for the color imbalance, additional gluons must be
exchanged. Although this can take place at low energy, it will still
suppress these diagrams somewhat, and so we do not include them
here. (Note: their inclusion does not change any of our conclusions.)

One important difference compared to two-body $B$-decay diagrams is
momentum dependence. In two-body decays, in the rest frame of the $B$,
the three-momenta of the final-state particles are equal and opposite.
One does not have the same type of behavior in three-body decays.
Although the sum of the three-momenta of the final particles is zero,
there is no constraint on any individual particle. As such, the
three-body diagrams are momentum dependent, and this must be taken
into account whenever the diagrams are used.

\section{Dalitz Plots}
\label{Dalitz}

In this section, we review certain aspects of the Dalitz-plot
analysis. To illustrate these, we focus on the decay $B^+ \to K^+
\pi^- \pi^+$ \cite{K+pi-pi+}. However, a similar type of analysis can
be applied to any three-body $B$ decay.

$B^+ \to K^+ \pi^- \pi^+$ can take place via intermediate resonances,
as well as non-resonant decays. The events in the Dalitz plot are
therefore described by the following two variables:
\bea
x &=& m^2_{K^+\pi^-} = \left( p_{K^+} + p_{\pi^-} \right)^2 ~, \nn\\
y &=& m^2_{\pi^+\pi^-} = \left( p_{\pi^+} + p_{\pi^-} \right)^2 ~.
\eea
Now, one of the great advantages of a Dalitz-plot analysis is that it
allows one to extract the full amplitude of the decay. To this end, we
write
\beq
{\cal M}(B^+ \to K^+ \pi^- \pi^+) = \sum_j c_j e^{i\theta_j} F_j(x,y) ~,
\label{Kpipiamp}
\eeq
where the sum is over all decay modes (resonant and non-resonant).
$c_j$ and $\theta_j$ are the magnitude and phase of the $j$
contribution, respectively, measured relative to one of the
contributing channels. The distributions $F_j$, which depend on $x$
and $y$, describe the dynamics of the individual decay amplitudes. In
the experimental analyses, these take different (known) forms for the
various contributions. The key point is that a maximum likelihood fit
over the entire Dalitz plot gives the best values of the $c_j$ and
$\theta_j$. Thus, the decay amplitude can be obtained.

In this paper, the following issue is of central importance. In $B^+
\to K^+ \pi^- \pi^+$, since the $\pi$'s are identical particles under
isospin, the overall $\pi^- \pi^+$ wavefunction must be symmetric. If
the $\pi\pi$ pair is in a state of even (odd) isospin, the
wavefunction (or, equivalently, the $B^+ \to K^+ \pi^- \pi^+$ decay
amplitude) must be symmetric (antisymmetric) under the exchange
$p_{\pi^+} \leftrightarrow p_{\pi^-}$. Unfortunately, the amplitude of
Eq.~(\ref{Kpipiamp}) does not possess such a symmetry.

It is the use of the parameters $x$ and $y$ which is problematic. A
better choice of variables would be $s_+$ and $s_-$, where
\bea
s_+ &=& m^2_{K^+\pi^+} = \left( p_{K^+} + p_{\pi^+} \right)^2 ~, \nn\\
x = s_- &=& m^2_{K^+\pi^-} = \left( p_{K^+} + p_{\pi^-} \right)^2 ~.
\eea
Now, under the exchange $p_{\pi^+} \leftrightarrow p_{\pi^-}$, we
simply have $s_+ \leftrightarrow s_-$. Thus, if we had started with the
amplitude ${\cal M}(B^+ \to K^+ \pi^- \pi^+) = g(s_+,s_-)$, the symmetric
combination would be $\frac{1}{\sqrt{2}}[g(s_+,s_-) + g(s_-,s_+)]$,
i.e.\ it would correspond to the production of the $\pi^- \pi^+$ pair
with a symmetric wavefunction; $\frac{1}{\sqrt{2}}[g(s_+,s_-) -
  g(s_-,s_+)]$ would be antisymmetric.

The problem is that the wavefunction of Eq.~(\ref{Kpipiamp}) is not
given in terms of $s_+$ and $s_-$. Fortunately, there is a resolution
to this problem: the independent Mandelstam variables $y$, $s_+$ and
$s_-$ satisfy
\beq
y = m_B^2 + 2m_\pi^2 + m_{K^+}^2 - s_+ - s_- ~.
\eeq
This implies that $f(x,y) = f(s_-,y) = f(s_-, m_B^2 + 2m_\pi^2 +
m_{K^+}^2 - s_+ - s_-) \equiv g(s_+,s_-)$. Given the decay amplitude
${\cal M}(x,y)$ of Eq.~(\ref{Kpipiamp}), one can therefore easily
construct the amplitude which is symmetric/antisymmetric in $p_{\pi^+}
\leftrightarrow p_{\pi^-}$. The same method applies to other $B \to
K\pi\pi$ decays, and indeed to all three-body decays. Thus, if there
are identical particles in the final state, the $B$-decay Dalitz plot
allows us to construct the amplitude for the production of these
particles in a symmetric/antisymmetric state.

Above, we argued that the Dalitz-plot analysis allows one to obtain
the amplitude ${\cal M}$ of any three-body $B$ decay. Actually, this
is not quite accurate -- the global phase of the amplitude is
undetermined. Thus, it is really $|{\cal M}|$ which should be compared
with theory. Similarly, one can obtain $|{\overline{\cal M}}|$ of the
CP-conjugate decay. In the rest of the paper, we refer to the
momentum-dependent branching ratio and direct CP asymmetry of a
particular decay. These are proportional to $|{\cal M}|^2 +
|{\overline{\cal M}}|^2$ and $|{\cal M}|^2 - |{\overline{\cal M}}|^2$,
respectively. Finally, for a self-conjugate final state such as
$K^0\pi^+\pi^-$ (where the $K^0$ is seen as $K_S$), the
momentum-dependent indirect CP asymmetry\footnote{The indirect CP
  asymmetry depends on the CP of the final state, and a-priori
  $K^0\pi^+\pi^-$ is a mixture of CP $+$ and CP $-$. However, the
  separation of symmetric and antisymmetric $\pi\pi$ states also fixes
  the final-state CP: $K^0(\pi\pi)_{sym}$ and $K^0(\pi\pi)_{anti}$
  have CP $+$ and $-$, respectively.} can be measured, and gives
${\cal M}^* {\overline{\cal M}}$ for this decay.

\section{\boldmath $B \to K\pi\pi$ Decays}

We begin with $B \to K\pi\pi$ decays, a $\btos$ transition. There are
six processes: $B^+ \to K^+\pi^+\pi^-$, $B^+ \to K^+\pi^0\pi^0$, $B^+
\to K^0\pi^+\pi^0$, $\bd \to K^+\pi^0\pi^-$, $\bd \to K^0\pi^+\pi^-$,
$\bd \to K^0\pi^0\pi^0$. In all of these, the overall wavefunction of
the final $\pi\pi$ pair must be symmetrized with respect to the
exchange of these two particles. There are two possibilities.  If the
relative angular momentum is even (odd), the isospin state must be
symmetric (antisymmetric). We refer to these two cases as
$I_{\pi\pi}^{sym}$ and $I_{\pi\pi}^{anti}$. As shown in
Sec.~\ref{Dalitz}, they can be determined experimentally.  We discuss
them in turn.

We first consider $I_{\pi\pi}^{sym}$, i.e.\ $I=(0,2)$. The final state
has $I=\frac12$, $\frac32$, or $\frac52$. The $B$-meson has
$I=\frac12$ and the weak Hamiltonian has $\Delta I = 0$ or 1. The
final state with $I=\frac52$ cannot be reached. So there are three
different ways of getting to the final state. Given that there are six
decays, this means that there should be three relations among their
amplitudes. This conclusion is an exact result; the relations can be
found by applying the Wigner-Eckart theorem:
\bea
\label{Kpipisymrels}
A(B^+ \to K^0\pi^+\pi^0)_{sym} &=& -A(\bd \to K^+\pi^0\pi^-)_{sym} ~, \\
\sqrt{2} A(B^+ \to K^0\pi^+\pi^0)_{sym} &=& A(\bd \to K^0\pi^+\pi^-)_{sym} + \sqrt{2} A(\bd \to K^0\pi^0\pi^0)_{sym} ~, \nn\\
\sqrt{2} A(\bd \to K^+\pi^0\pi^-)_{sym} &=& A(B^+ \to K^+\pi^+\pi^-)_{sym} + \sqrt{2} A(B^+ \to K^+\pi^0\pi^0)_{sym} ~. \nn
\eea
These relations were first given (implicitly) in Ref.~\cite{LNQS}. The
subscript `$sym$' indicates that the $\pi\pi$ isospin state is
symmetrized.

In terms of diagrams, the amplitudes are given by
\bea
\sqrt{2} A(B^+ \to K^0\pi^+\pi^0)_{sym} &=& -T'_1 e^{i\gamma}-C'_2 e^{i\gamma} + P'_{EW2} + P^{\prime C}_{EW1} ~, \nn\\
A(\bd \to K^0\pi^+\pi^-)_{sym} &=& -T'_1 e^{i\gamma}-C'_1 e^{i\gamma}-{\tilde P}'_{uc} e^{i\gamma}+ {\tilde P}'_{tc} \nn\\
&& \hskip1.5truecm +~\frac13 P'_{EW1} + \frac23 P^{\prime C}_{EW1} - \frac13 P^{\prime C}_{EW2} ~, \nn\\
\sqrt{2} A(\bd \to K^0\pi^0\pi^0)_{sym} &=& C'_1 e^{i\gamma}- C'_2 e^{i\gamma}+{\tilde P}'_{uc} e^{i\gamma}- {\tilde P}'_{tc}  \nn\\
&& \hskip1.5truecm -~\frac13 P'_{EW1} + P'_{EW2} +\frac13 P^{\prime C}_{EW1} + \frac13 P^{\prime C}_{EW2} ~, \nn\\
A(B^+ \to K^+\pi^+\pi^-)_{sym} &=& -T'_2 e^{i\gamma}-C'_1 e^{i\gamma}-{\tilde P}'_{uc} e^{i\gamma}+ {\tilde P}'_{tc} \nn\\
&& \hskip1.5truecm +~\frac13 P'_{EW1} - \frac13 P^{\prime C}_{EW1} + \frac23 P^{\prime C}_{EW2} ~, \nn\\
\sqrt{2} A(B^+ \to K^+\pi^0\pi^0)_{sym} &=& T'_1 e^{i\gamma}+T'_2 e^{i\gamma}+C'_1 e^{i\gamma}+C'_2 e^{i\gamma}+{\tilde P}'_{uc} e^{i\gamma}- {\tilde P}'_{tc} \nn\\
&& \hskip1.5truecm -~\frac13 P'_{EW1} - P'_{EW2} - \frac23 P^{\prime C}_{EW1} - \frac23 P^{\prime C}_{EW2} ~, \nn\\
\sqrt{2} A(\bd \to K^+\pi^0\pi^-)_{sym} &=& T'_1 e^{i\gamma}+C'_2 e^{i\gamma}- P'_{EW2} - P^{\prime C}_{EW1} ~, 
\label{Kpipisymamps}
\eea
where ${\tilde P}' \equiv P'_1 +P'_2$. (Note: all amplitudes have been
multiplied by $\sqrt{2}$.)  Above we have explicitly written the
weak-phase dependence (including the minus sign from $V_{tb}^*
V_{ts}$ [${\tilde P}'_{tc}$ and EWP's]), while the diagrams contain strong
phases. (The phase information in the Cabibbo-Kobayashi-Maskawa quark
mixing matrix is conventionally parametrized in terms of the unitarity
triangle, in which the interior (CP-violating) angles are known as
$\alpha$, $\beta$ and $\gamma$ \cite{pdg}.)  It is straightforward to
verify that the three relations of Eq.~(\ref{Kpipisymrels}) are
reproduced. Thus, in this case, there is no difference between the
exact and diagrammatic amplitude relations.

We now turn to $I_{\pi\pi}^{anti}$, i.e.\ $I=1$. Here there are four
processes: $B^+ \to K^+\pi^+\pi^-$, $B^+ \to K^0\pi^+\pi^0$, $\bd \to
K^+\pi^0\pi^-$, $\bd \to K^0\pi^+\pi^-$ (one cannot antisymmetrize a
$\pi^0\pi^0$ state).  The final state has $I=\frac12$ or $\frac32$, so
there are still three different paths to get to the final state. We
therefore expect one relation among the four
amplitudes. Ref.~\cite{LNQS} notes that it is similar to that in
$B\to\pi K$:
\bea
&& \sqrt{2} A(B^+ \to K^+\pi^+\pi^-)_{anti} + A(B^+ \to K^0\pi^+\pi^0)_{anti} = \nn\\
&& \hskip1.5truecm \sqrt{2} A(\bd \to K^0\pi^+\pi^-)_{anti} + A(\bd \to K^+\pi^0\pi^-)_{anti} ~,
\label{Kpipiantirel}
\eea
where the subscript `$anti$' indicates that the $\pi\pi$ isospin
state is antisymmetrized.

Writing the amplitudes in terms of diagrams is a bit more complicated
because antisymmetrization is involved. Depending on the order of the
pions, there might be an extra minus sign. To account for this, we
use the following prescription:
\begin{itemize}

\item All diagrams with the pions in order of decreasing charge from
  top to bottom are unmodified; all diagrams with the pions in order
  of increasing charge from top to bottom get an additional factor of
  $-1$.

\end{itemize}
This requires that diagrams always be drawn the same way. For example,
the spectator quark for all tree diagrams should always appear in the
same place (e.g.\ at the bottom of the diagram), and the decay
products of the neutral bosons in penguin and EWP diagrams should
always appear in the same order (e.g.\ quark on top, antiquark on the
bottom).

With this rule, the amplitudes take the form\footnote{Note: even
  though the diagrams of Eq.~(\ref{Kpipiantisymamps}) have the same
  names as those of Eq.~(\ref{Kpipisymamps}), they are not the same
  diagrams. That is, in general, they take different values.}
\bea
\sqrt{2} A(B^+ \to K^0\pi^+\pi^0)_{anti} &=& -T'_1 e^{i\gamma}-C'_2 e^{i\gamma}-2 {\tilde P}'_{uc} e^{i\gamma}
+2 {\tilde P}'_{tc} \nn\\
&& \hskip1.5truecm -~P'_{EW2} - \frac13 P^{\prime C}_{EW1} + \frac23 P^{\prime C}_{EW2} ~, \nn\\
A(\bd \to K^0\pi^+\pi^-)_{anti} &=& -T'_1 e^{i\gamma}-C'_1 e^{i\gamma}-{\tilde P}'_{uc} e^{i\gamma}+ {\tilde
  P}'_{tc} \nn\\
&& \hskip1.5truecm +~P'_{EW1} - \frac23 P^{\prime C}_{EW1} + \frac13 P^{\prime C}_{EW2}
~, \nn\\
A(B^+ \to K^+\pi^+\pi^-)_{anti} &=& T'_2 e^{i\gamma}-C'_1 e^{i\gamma} +{\tilde P}'_{uc} e^{i\gamma}- {\tilde
  P}'_{tc} \nn\\
&& \hskip1.5truecm +~P'_{EW1} - \frac13 P^{\prime C}_{EW1} + \frac23 P^{\prime C}_{EW2}
~, \nn\\
\sqrt{2} A(\bd \to K^+\pi^0\pi^-)_{anti} &=& T'_1 e^{i\gamma} +2 T'_2 e^{i\gamma}-C'_2 e^{i\gamma}+2 {\tilde
  P}'_{uc} e^{i\gamma}-2 {\tilde P}'_{tc} \nn\\
&& \hskip1.5truecm -~P'_{EW2} +\frac13 P^{\prime C}_{EW1} + \frac43 P^{\prime
  C}_{EW2} ~.
\label{Kpipiantisymamps}
\eea
(As above, all amplitudes have been multiplied by $\sqrt{2}$.)  The
relation of Eq.~(\ref{Kpipiantirel}) is reproduced. Therefore, there
is no difference between the exact and diagrammatic amplitude
relations in the antisymmetric case.

\subsection{Resonances}

It is possible that the $B$ decays to an intermediate on-shell $M_1
M_2$ state, which then subsequently decays to $K\pi\pi$. Examples of
such resonances are $M_1 M_2 = K\rho$, $K^* \pi$, $K f_0(980)$. The
question now is: how does the diagrammatic analysis presented above
jibe with resonant decays? To answer this, we examine the resonances
in turn.

Consider first $M_1 M_2 = K\rho$. The four decays are $B^+ \to
K^+\rho^0$, $B^+ \to K^0\rho^+$, $\bd \to K^0\rho^0$, $\bd \to
K^+\rho^-$, whose amplitudes take the form
\bea
\sqrt{2} A(B^+ \to K^+\rho^0) &=& - T'_V e^{i\gamma} - C'_P
e^{i\gamma} - P'_{uc,V} e^{i\gamma} + P'_{tc,V} + P'_{EW,P} + \frac23
P_{EW,V}^{\prime C} ~, \nn\\
A(B^+ \to K^0\rho^+) &=& P'_{uc,V} e^{i\gamma} - P'_{tc,V} + \frac13
P_{EW,V}^{\prime C} ~, \nn\\
\sqrt{2} A(\bd \to K^0\rho^0) &=& - C'_P e^{i\gamma} + P'_{uc,V}
e^{i\gamma} - P'_{tc,V} + P'_{EW,P} + \frac13 P_{EW,V}^{\prime C} ~,
\nn\\
A(\bd \to K^+\rho^-) &=& - T'_V e^{i\gamma} - P'_{uc,V} e^{i\gamma} +
P'_{tc,V} + \frac23 P_{EW,V}^{\prime C} ~,
\eea
where the subscript $P$ or $V$ indicates which final-state meson
[pseudoscalar ($K$) or vector ($\rho$)] contains the spectator quark
of the $B$ meson \cite{PV}. (Note that the diagrams which describe
resonant decays are a subset of those used for $B \to K\pi\pi$
(Fig.~\ref{BPPPfig}). Above, the diagram $D_V$ ($D_P$) is the same as
$D_2$ ($D_1$).) The relation among the amplitudes is
\bea
&& \sqrt{2} A(B^+ \to K^+\rho^0) + A(B^+ \to K^0\rho^+) = \nn\\
&& \hskip1.5truecm \sqrt{2} A(\bd \to K^0\rho^0) + A(\bd \to K^+\rho^-) ~.
\eea
Given that $\rho^0 \to \pi^+\pi^-$, $\rho^+ \to \pi^+\pi^0$ and
$\rho^- \to \pi^0\pi^-$, this reproduces Eq.~(\ref{Kpipiantirel}),
which is the relation for the antisymmetric $\pi\pi$ isospin
state. This makes sense, since the $\rho$ decays to $(\pi\pi)_{anti}$.

Consider now $M_1 M_2 = K f_0(980)$. There are two decays: $B^+ \to
K^+ f_0(980)$ and $\bd \to K^0 f_0(980)$. It is straightforward to show
that there is no relation between the two amplitudes. However, the
$f_0(980)$ decays to a pion pair in a symmetric isospin state, with
$A(f_0 \to (\pi^+\pi^-)_{sym}) = -\sqrt{2} A(f_0 \to \pi^0\pi^0)$. This leads to 
\bea
A(\bd \to K^0\pi^+\pi^-) + \sqrt{2} A(\bd \to K^0\pi^0\pi^0) &=& 0 ~, \nn\\
A(B^+ \to K^+\pi^+\pi^-) + \sqrt{2} A(B^+ \to K^+\pi^0\pi^0) &=& 0 ~.
\eea
Given that the $K f_0(980)$ resonance does not contribute to $A(B^+
\to K^0\pi^+\pi^0)$, $A(\bd \to K^+\pi^0\pi^-)$ or $A(B^+ \to
K^0\pi^+\pi^0)$, the decays $B \to K f_0(980) \to K\pi\pi$ satisfy
Eq.~(\ref{Kpipisymrels}), which are the relations for the symmetric
$\pi\pi$ isospin state.

Finally, consider $M_1 M_2 = K^* \pi$.  The four decays are $B^+ \to
K^{*0}\pi^+$, $B^+ \to K^{*+} \pi^0$, $\bd \to K^{*+}\pi^-$, $\bd \to
K^{*0}\pi^0$. The amplitudes are \cite{PV}
\bea
A(B^+ \to K^{*0}\pi^+) &=& P'_{uc,P} e^{i\gamma} - P'_{tc,P} + \frac13
P_{EW,P}^{\prime C} ~, \nn\\
\sqrt{2} A(B^+ \to K^{*+}\pi^0) &=& - T'_P e^{i\gamma} - C'_V
e^{i\gamma} - P'_{uc,P} e^{i\gamma} + P'_{tc,P} + P'_{EW,V} + \frac23
P_{EW,P}^{\prime C} ~, \nn\\
A(\bd \to K^{*+}\pi^-) &=& - T'_P e^{i\gamma} - P'_{uc,P} e^{i\gamma} +
P'_{tc,P} + \frac23 P_{EW,P}^{\prime C} ~, \nn\\
\sqrt{2} A(\bd \to K^{*0}\pi^0) &=& - C'_V e^{i\gamma} + P'_{uc,P}
e^{i\gamma} - P'_{tc,P} + P'_{EW,V} + \frac13 P_{EW,P}^{\prime C} ~.
\eea
The relation among the amplitudes is
\bea
&& A(B^+ \to K^{*0}\pi^+) + \sqrt{2} A(B^+ \to K^{*+}\pi^0) = \nn\\
&& \hskip1.5truecm A(\bd \to K^{*+}\pi^-) + \sqrt{2} A(\bd \to K^{*0}\pi^0) ~.
\label{K*pirel}
\eea
Now, the $K^*$ decays to $K\pi$, and both charge assignments are allowed:
\bea
K^{*+} &\to& \sqrt{1/3} \, K^+ \pi^0 - \sqrt{2/3} \, K^0 \pi^+ ~, \nn\\
K^{*0} &\to& \sqrt{2/3} \, K^+ \pi^- - \sqrt{1/3} \, K^0 \pi^0 ~.
\label{K*decay}
\eea
There are therefore several $K^* \pi$ contributions to a particular
$K\pi\pi$ final state. However, one never reproduces the relations in
Eqs.~(\ref{Kpipisymrels}) or (\ref{Kpipiantirel}). This reflects the
fact that this resonance contributes to both $(\pi\pi)_{sym}$ and
$(\pi\pi)_{anti}$. 

Still, it is instructive to examine the relation obtained when the
resonance decays. This is obtained by inserting Eq.~(\ref{K*decay})
into Eq.~(\ref{K*pirel}). When the $\pi\pi$ pair is in a symmetric
isospin state, one has
\bea
&& \sqrt{2} A(B^+ \to K^+\pi^+\pi^-) - 3 A(B^+ \to K^0\pi^+\pi^0) + 
\sqrt{2} A(B^+ \to K^+\pi^0\pi^0) = \nn\\
&& \hskip0.5truecm 3 A(\bd \to K^+\pi^0\pi^-) - \sqrt{2} A(\bd \to
K^0\pi^+\pi^-) - \sqrt{2} A(\bd \to K^0\pi^0\pi^0) ~.
\eea
This is obviously not the same as Eq.~(\ref{Kpipisymrels}). This is
because there are only four $B \to K^* \pi$ decays (and not six, as in
$B \to K\pi\pi$), and so there is only one relation among the
$K\pi\pi$ decays.

On the other hand, the case where the $\pi\pi$ pair is in an
antisymmetric isospin state is more interesting. For
$I_{\pi\pi}^{anti}$, amplitudes to final states with two $\pi^0$'s are
zero. Also, there is an additional factor of $-1$ if the pions are in
order of increasing charge from top to bottom. Taking the $K^*$ in $B
\to K^* \pi$ to be on top of the $\pi$, the amplitudes $A(B^+ \to
K^+\pi^+\pi^-)_{K^{*0}\pi^+}$, $A(B^+ \to
K^0\pi^+\pi^0)_{K^{*0}\pi^+}$ and $A(\bd \to
K^+\pi^0\pi^-)_{K^{*0}\pi^0}$ all get an extra minus sign (the
subscript indicates the resonance which gives rise to the final
state). When these are taken into account, the insertion of
Eq.~(\ref{K*decay}) into Eq.~(\ref{K*pirel}) gives the relation in
Eq.~(\ref{Kpipiantirel}). We therefore see that the $B\to K\pi\pi$
amplitude relation is reproduced by $B \to K^* \pi$ decays for the
$I_{\pi\pi}^{anti}$ case.

The point here is that it is useful to consider the entire $B \to M_1
M_2 \to K\pi\pi$ decay chain, and that the distinction between
$I_{\pi\pi}^{sym}$ and $I_{\pi\pi}^{anti}$ is important, even for
resonances.

\subsection{Penguin Dominance}

In general, the dominant contribution to $\btos$ transitions comes
from the penguin amplitude. In Ref.~\cite{GR2005}, Gronau and Rosner
explore the consequences for $B \to K\pi\pi$ decays of assuming
penguin dominance and neglecting all other contributions. They note
that, in this limit, the amplitudes must respect isospin reflection
(i.e.\ $u \leftrightarrow d$), which implies that
\bea
A(B^+ \to K^+\pi^+\pi^-) &=& A(\bd \to K^0\pi^+\pi^-) ~, \nn\\
A(B^+ \to K^0\pi^+\pi^0) &=& A(\bd \to K^+\pi^0\pi^-) ~, \nn\\
A(\bd \to K^0\pi^0\pi^0) &=& A(B^+ \to K^+\pi^0\pi^0) ~,
\eea
up to possible relative signs. They find that, on the whole, the data respect
these relations.

The expression of the amplitudes in terms of diagrams allows us to go
beyond these results. Using the method of Sec.~\ref{Dalitz} to
distinguish $I_{\pi\pi}^{sym}$ and $I_{\pi\pi}^{anti}$, it is possible
to consider the two cases separately, under the condition that only
the diagram ${\tilde P}'_{tc}$ is retained in the amplitudes.

In the symmetric scenario, we have the following predictions:
\bea
A(B^+ \to K^0\pi^+\pi^0) &=& A(\bd \to K^+\pi^0\pi^-) = 0 ~, \nn\\
&& \hskip-5truecm A(B^+ \to K^+\pi^+\pi^-) = A(\bd \to K^0\pi^+\pi^-) \nn\\
&& \hskip-1.8truecm =~-\sqrt{2} A(\bd \to K^0\pi^0\pi^0) = -\sqrt{2} A(B^+ \to K^+\pi^0\pi^0) ~.
\eea
And in the antisymmetric scenario, we have
\bea
A(\bd \to K^0\pi^0\pi^0) &=& A(B^+ \to K^+\pi^0\pi^0) = 0 ~, \nn\\
&& \hskip-5truecm A(B^+ \to K^0\pi^+\pi^0) = -A(\bd \to K^+\pi^0\pi^-) \nn\\ 
&& \hskip-1.8truecm =~- \sqrt{2} A(B^+ \to K^+\pi^+\pi^-) = \sqrt{2} A(\bd \to K^0\pi^+\pi^-) ~.
\eea
These provide further tests of the SM.

In fact, several of these decays have been measured: $B^+ \to
K^+\pi^+\pi^-$ \cite{K+pi-pi+,K+pi-pi+new}, $\bd \to K^0\pi^+\pi^-$
\cite{K0pi-pi+}, and $\bd \to K^+\pi^0\pi^-$ \cite{K+pi0pi-}. We can
therefore test some of the above relations. Specifically, in terms of
branching ratios (integrated over the entire Dalitz plot), the
predictions are
\bea
\mathcal{B}(K^+ \pi^0 \pi^-)_{sym} & = & 0 ~, \nn\\
\mathcal{B}(K^+ \pi^+ \pi^-)_{sym} & = & \left( \tau_+ /\tau_0 \right) \mathcal{B}(K^0 \pi^+ \pi^-)_{sym} ~, \nn\\
\frac12 \left( \tau_+ /\tau_0 \right) \mathcal{B}(K^+ \pi^0 \pi^-)_{anti} & = & \mathcal{B}(K^+ \pi^+ \pi^-)_{anti} = 
\left( \tau_+ /\tau_0 \right) \mathcal{B}(K^0 \pi^+ \pi^-)_{anti} ~.
\label{predictions}
\eea

We determine the symmetric and antisymmetric amplitudes for the three
decays using the Dalitz-plot method described in
Sec.~\ref{Dalitz}. Consider first $B^+ \to K^+\pi^+\pi^-$. We write
this amplitude in terms of $x \equiv (p_{K^+} + p_{\pi^+})^2$ and $y
\equiv (p_{K^+} + p_{\pi^-})^2$. Given the decay amplitude $f(x,y)$,
the symmetric amplitude is taken to be $f_{sym} = \frac{1}{\sqrt{2}}
(f(x,y) + f(y,x))$, and we compute the integral of $|f_{sym}|^2$ and
$|f|^2$ over the Dalitz plot\footnote{Note that, because of the
  coefficient $\frac{1}{\sqrt{2}}$ in $f_{sym}$, one must integrate
  over only half of the Dalitz plot to avoid double
  counting. Alternatively, $f_{sym}$ can be defined with a factor
  $\frac12$, and one integrates over the entire Dalitz plot. There are
  no such issues with $f$.}. A similar procedure is carried out for
the antisymmetric amplitude $f_{anti} = \frac{1}{\sqrt{2}} (f(x,y) -
f(y,x))$. The other two decays are treated in the same way.

Although the full amplitudes for $B^+ \to K^+\pi^+\pi^-$ and $\bd \to
K^0\pi^+\pi^-$ are split roughly equally between symmetric and
antisymmetric, the same is not true for $\bd \to K^+\pi^0\pi^-$:
\bea
\Gamma (K^+ \pi^+ \pi^-)_{sym} = 0.65 \, \Gamma (K^+ \pi^+ \pi^-) ~, \nn\\
\Gamma (K^0 \pi^+ \pi^-)_{sym} = 0.68 \, \Gamma (K^0 \pi^+ \pi^-) ~, \nn\\
\Gamma (K^+ \pi^0 \pi^-)_{sym} = 0.11 \, \Gamma (K^+ \pi^0 \pi^-) ~.
\eea
With these, we obtain
\bea
\mathcal{B}(K^+ \pi^0 \pi^-)_{sym} & = & (4.0 \pm 0.3) \times 10^{-6} ~, \nn\\
\mathcal{B}(K^+ \pi^+ \pi^-)_{sym} & = & (33.3 \pm 2.0) \times 10^{-6} ~, \nn\\
\left( \tau_+ /\tau_0 \right) \mathcal{B}(K^0 \pi^+ \pi^-)_{sym} & = & (36.4 \pm 1.5) \times 10^{-6} ~, \nn\\
\frac12 \left( \tau_+ /\tau_0 \right) \mathcal{B}(K^+ \pi^0 \pi^-)_{anti} & = & (17.1 \pm 1.3) \times 10^{-6} ~, \nn\\
\mathcal{B}(K^+ \pi^+ \pi^-)_{anti} & = & (17.6 \pm 1.0) \times 10^{-6} ~, \nn\\
\left( \tau_+ /\tau_0 \right) \mathcal{B}(K^0 \pi^+ \pi^-)_{anti} & = & (17.0 \pm 0.7) \times 10^{-6} ~.
\eea
(Note that the above errors do not include the errors in the
parameters obtained from the Dalitz-plot analyses of the three
decays.) We therefore see that the data agree with the predictions of
Eq.~(\ref{predictions}). In particular, $\mathcal{B}(K^+ \pi^0
\pi^-)_{sym}$ is indeed greatly suppressed, in agreement with the SM.

\subsection{Weak-Phase Information}
\label{Kpipiweak}

Since the expressions for the decay amplitudes include the weak phase
$\gamma$, it is natural to ask whether $\gamma$ can be extracted from
measurements of $B \to K\pi\pi$ decays. The answer is `yes' if the
number of unknown theoretical parameters in the amplitudes is less
than or equal to the number of observables. In performing this
comparison, we examine separately the $I_{\pi\pi}^{sym}$ and
$I_{\pi\pi}^{anti}$ scenarios.

Consider first the $I_{\pi\pi}^{sym}$ case. Here there are six $B \to
K\pi\pi$ decays. On the other hand, the first relation in
Eq.~(\ref{Kpipisymrels}) shows that the amplitudes for $B^+ \to
K^0\pi^+\pi^0$ and $\bd \to K^+\pi^0\pi^-$ are equal (up to a sign),
so that there are only five independent decays. The Dalitz-plot
analyses of these decays allow one to obtain the momentum-dependent
branching ratios and direct CP asymmetries of $B^+ \to K^+\pi^+\pi^-$,
$B^+ \to K^+\pi^0\pi^0$, $\bd \to K^+\pi^0\pi^-$, $\bd \to
K^0\pi^+\pi^-$, and $\bd \to K^0\pi^0\pi^0$.  In addition, one can
measure the momentum-dependent indirect CP asymmetry of $\bd \to
K^0\pi^+\pi^-$. (The indirect CP asymmetry of $\bd \to K^0\pi^0\pi^0$
will be very difficult, if not impossible, to measure.) Thus, there
are essentially 11 (momentum-dependent) observables in
$I_{\pi\pi}^{sym}$ $B \to K\pi\pi$ decays.

For the case of $I_{\pi\pi}^{anti}$, there are four decays, yielding 9
observables: the momentum-dependent branching ratios and direct CP
asymmetries of $B^+ \to K^+\pi^+\pi^-$, $B^+ \to K^0\pi^+\pi^0$, $\bd
\to K^+\pi^0\pi^-$, $\bd \to K^0\pi^+\pi^-$, and the
momentum-dependent indirect CP asymmetry of $\bd \to
K^0\pi^+\pi^-$. Since this is fewer than above, we conclude that the
$I_{\pi\pi}^{sym}$ scenario is the more promising for extracting
$\gamma$.

The six $I_{\pi\pi}^{sym}$ amplitudes are given in
Eq.~(\ref{Kpipisymamps}). Although there are a large number of
diagrams in these amplitudes, they can be combined into a smaller
number of effective diagrams:
\bea
\sqrt{2} A(B^+ \to K^0\pi^+\pi^0)_{sym} &=& - T'_a e^{i\gamma} - T'_b e^{i\gamma} + P'_{EW,a} + P'_{EW,b} ~, \nn\\
A(\bd \to K^0\pi^+\pi^-)_{sym} &=& - T'_a e^{i\gamma} - P'_a e^{i\gamma} + P'_b ~, \nn\\
\sqrt{2} A(\bd \to K^0\pi^0\pi^0)_{sym} &=& - T'_b e^{i\gamma} + P'_a e^{i\gamma} - P'_b + P'_{EW,a} + P'_{EW,b} ~, \nn\\
A(B^+ \to K^+\pi^+\pi^-)_{sym} &=& - P'_a e^{i\gamma} + P'_b - P'_{EW,a} ~, \nn\\
\sqrt{2} A(B^+ \to K^+\pi^0\pi^0)_{sym} &=& T'_a e^{i\gamma} + T'_b e^{i\gamma} + P'_a e^{i\gamma} - P'_b - P'_{EW,b} ~, \nn\\
\sqrt{2} A(\bd \to K^+\pi^0\pi^-)_{sym} &=& T'_a e^{i\gamma} + T'_b e^{i\gamma} - P'_{EW,a} - P'_{EW,b} ~,
\label{Kpipieffamps}
\eea
where
\bea
T'_a &\equiv& T'_1 - T'_2 ~,\nn\\
T'_b &\equiv& C'_2 + T'_2 ~,\nn\\
P'_a &\equiv& {\tilde P}'_{uc} + T'_2 + C'_1 ~,\nn\\
P'_b &\equiv& {\tilde P}'_{tc} + \frac13 P'_{EW1} + \frac23 P^{\prime C}_{EW1} - \frac13 P^{\prime C}_{EW2} ~,  \nn\\
P'_{EW,a} &\equiv& P ^{\prime C}_{EW1} - P^{\prime C}_{EW2} ~, \nn\\
P'_{EW,b} &\equiv& P'_{EW2} + P^{\prime C}_{EW2} ~.
\label{eq:effdiag}
\eea
The amplitudes can therefore be written in terms of 6 effective
diagrams. This corresponds to 12 theoretical parameters\footnote{In
  fact, there is another theoretical parameter -- the phase of
  $\bd$-$\bdbar$ mixing, $\beta$, enters in the expression for the
  indirect CP asymmetry. However, the value for $\beta$ can be taken
  from the indirect CP asymmetry in $\bd\to J/\psi \ks$ \cite{pdg}.}:
6 magnitudes of diagrams, 5 relative (strong) phases, and $\gamma$. We
remind the reader that the diagrams are momentum dependent. This does
not pose a problem. They will be determined via a fit to the data. But
since the experimental observables are themselves momentum dependent,
the fit will yield the momentum dependence of each diagram. 

Unfortunately, as noted above, there are only 11 experimental
observables. Therefore, in order to extract weak-phase information
($\gamma$), one requires additional input.

A previous analysis made an attempt in this direction. In 2003,
Deshpande, Sinha and Sinha (DSS) wrote schematic expressions for the
symmetric $B \to K\pi\pi$ amplitudes, including tree and EWP
contributions \cite{DSS}. Now, in $B\to\pi K$ decays, it was shown
that, under flavor SU(3) symmetry, the EWP diagrams are proportional
to the tree diagrams (apart from their weak phases) \cite{EWPs}. DSS
assumed that the EWP and tree contributions to $B^+ \to K^0\pi^+\pi^0$
are related in the same way. This gives the additional input, and
allows the measurement of $\gamma$.  Unfortunately, it was
subsequently noted that the assumed EWP-tree relation in $K\pi\pi$
does not hold \cite{Grocomment}, so that $\gamma$ cannot be
extracted. This is the present situation.

In fact, the situation can be remedied. Referring to the $\bd \to
K^0\pi^+\pi^0$ amplitude in Eq.~(\ref{Kpipisymamps}), DSS made the
assumption that $T'_1 + C'_2$ is related to $P'_{EW2} + P^{\prime
  C}_{EW1}$, and this was shown not to be true. We agree with this.
However, there are other EWP-tree relations which do hold, and their
inclusion does allow the extraction of $\gamma$. The full derivation
is rather complicated, and so we present this in a separate paper
\cite{Kpipigamma}.

Finally, we note that there is another method for obtaining $\gamma$
from $B \to K\pi\pi$ decays.  In two-body $\btos$ $B$ decays, the
diagrams are expected to obey the approximate hierarchy \cite{GHLR}
\bea
1 &:& P'_{tc} ~, \nn\\
{\bar\lambda} &:& T', P'_{EW} ~, \nn\\
{\bar\lambda}^2 &:& C', P'_{uc}, P^{\prime C}_{EW}  ~,
\label{hierarchy}
\eea
where ${\bar\lambda} \simeq 0.2$. If the three-body decay diagrams
obey a similar hierarchy, one can neglect $C'_1$, $C'_2$, ${\tilde
  P}'_{uc}$, $P^{\prime C}_{EW1}$, $P^{\prime C}_{EW2}$, and incur
only a $\sim 5\%$ theoretical error. But if these diagrams are
neglected, then two of the effective diagrams vanish: $P'_{EW,a} \to
0$ and $T'_b - P'_a \to 0$ [Eq.~(\ref{eq:effdiag})]. In this case, the
amplitudes can be written in terms of 4 effective diagrams,
corresponding to 8 theoretical parameters: 4 magnitudes of diagrams, 3
relative (strong) phases, and $\gamma$. Given that there are 11
experimental observables, the weak phase $\gamma$ can be
extracted\footnote{This technique does not work when the $\pi\pi$ pair
  is in an antisymmetric state of isospin. In this case, there are
  still more theoretical unknowns than observables, so that $\gamma$
  cannot be extracted.}.

The downside of this method is that it is difficult to test the
assumption that certain diagrams are negligible. Indeed, the presence
of resonances may change the hierarchy. In light of this, the
theoretical error is uncertain, and this must be addressed if this
method is used.

\section{\boldmath $B \to KK{\bar K}$ Decays}

We now turn to $B \to KK{\bar K}$ decays, also a $\btos$
transition. The four processes are: $B^+ \to K^+ K^+ K^-$, $B^+ \to
K^+ K^0 \kbar$, $\bd \to K^+ K^0 K^-$, $\bd \to K^0 K^0 \kbar$.
Here the overall wavefunction of the final $KK$ pair must be
symmetrized. If the relative angular momentum is even, the isospin
state must be symmetric ($I=1$); if it is odd, the isospin state must
be antisymmetric ($I=0$).

For the symmetric case, the final state has $I=\frac12$ or $\frac32$,
so there are three different ways of reaching it. There should
therefore be one relation among the four decay amplitudes. From the
Wigner-Eckart theorem, it is
\bea
&& A(B^+ \to K^+ K^+ K^-)_{sym} + \sqrt{2} A(B^+ \to K^+ K^0 \kbar)_{sym} = \nn\\
&& \hskip1.5truecm \sqrt{2} A(\bd \to K^+ K^0 K^-)_{sym} + A(\bd \to K^0 K^0 \kbar)_{sym} ~.
\label{KKKrel}
\eea

In terms of diagrams, the amplitudes are given by
\bea
\label{KKKsym}
A(B^+ \to K^+ K^+ K^-)_{sym} &=& -T'_{2,s} e^{i\gamma}-C'_{1,s} e^{i\gamma}
-{\hat P}'_{uc} e^{i\gamma}+ {\hat P}'_{tc} \nn\\
&& \hskip0.8truecm +~\frac23 P'_{EW1,s} - \frac13 P'_{EW1} + \frac23
P^{\prime C}_{EW2,s} - \frac13 P^{\prime C}_{EW1} ~, \nn\\
\sqrt{2} A(B^+ \to K^+ K^0 \kbar)_{sym} &=& {\hat P}'_{uc} e^{i\gamma}- {\hat P}'_{tc} \nn\\
&& \hskip0.8truecm +~\frac13 P'_{EW1,s} + \frac13 P'_{EW1} + \frac13
P^{\prime C}_{EW2,s} + \frac13 P^{\prime C}_{EW1} ~, \nn\\
\sqrt{2} A(\bd \to K^+ K^0 K^-)_{sym} &=& -T'_{2,s} e^{i\gamma}-C'_{1,s} e^{i\gamma}
-{\hat P}'_{uc} e^{i\gamma}+ {\hat P}'_{tc} \\
&& \hskip0.8truecm +~\frac23 P'_{EW1,s} - \frac13 P'_{EW1} + \frac23
P^{\prime C}_{EW2,s} - \frac13 P^{\prime C}_{EW1} ~, \nn\\
A(\bd \to K^0 K^0 \kbar)_{sym} &=& {\hat P}'_{uc} e^{i\gamma}- {\hat P}'_{tc} \nn\\
&& \hskip0.8truecm +~\frac13 P'_{EW1,s} + \frac13 P'_{EW1} + \frac13
P^{\prime C}_{EW2,s} + \frac13 P^{\prime C}_{EW1} ~, \nn
\eea
where ${\hat P}' \equiv P'_{2,s} + P'_1$. It is straightforward to
verify that the relation of Eq.~(\ref{KKKrel}) is reproduced. On the
other hand, one sees that there are, in fact, two relations:
\bea
A(B^+ \to K^+ K^+ K^-)_{sym} &=& \sqrt{2} A(\bd \to K^+ K^0 K^-)_{sym} ~, \nn\\
\sqrt{2} A(B^+ \to K^+ K^0 \kbar)_{sym} &=& A(\bd \to K^0 K^0 \kbar)_{sym} ~.
\label{KKKapproxrels}
\eea
What's happening is the following. Eq.~(\ref{KKKrel}) is
exact. However, when annihilation-type diagrams are neglected -- as is
done in our diagrammatic expressions of amplitudes -- then one finds
the two relations above. This is an example of how one can go beyond
the exact relations if certain negligible diagrams are dropped. 

In order to test these relations, it is necessary to isolate the
symmetric piece of the decay amplitudes. $B^+ \to K^+ K^+ K^-$ and
$\bd \to K^0 K^0 \kbar$ are automatically symmetric since the final
states contain truly identical particles. On the other hand, for $\bd
\to K^+ K^0 K^-$ and $B^+ \to K^+ K^0 \kbar$, the symmetric amplitude
can be obtained using the Dalitz-plot method of Sec.~\ref{Dalitz}.
Now, the Dalitz plot of $\bd \to K^+ K^0 K^-$ has already been
measured \cite{KKKBelle, KKKBabar}. This allows us to test the first
relation in Eq.~(\ref{KKKapproxrels}).

We use the Dalitz-plot analysis of $\bd \to K^+ K_S K^-$ given in
Ref.~\cite{KKKBelle}, with $A(\bd \to K^+ K^0 K^-) = \sqrt{2} A(\bd
\to K^+ K_S K^-)$. We find $\Gamma (\bd \to K^+ K^0 K^-)_{sym} = 0.57
\, \Gamma (\bd \to K^+ K^0 K^-)$. This then gives
\beq
2 \left( \tau_+ /\tau_0 \right) \mathcal{B} (\bd \to K^+ K^0 K^-)_{sym} = (30.0 \pm 2.8) \times 10^{-6} ~.
\eeq
(Note that the above error does not include the errors in the
parameters obtained from the Dalitz-plot analysis of
Ref.~\cite{KKKBelle}.) This is to be compared with \cite{hfag}
\beq
\mathcal{B} (B^+ \to K^+ K^+ K^-) = (32.5 \pm 1.5) \times 10^{-6} ~.
\eeq
We therefore see that the first relation in Eq.~(\ref{KKKapproxrels})
is satisfied. This supports our assumption that annihilation-type
diagrams are negligible.

In the antisymmetric case, there are only two decays: $B^+ \to K^+ K^0
\kbar$ and $\bd \to K^+ K^0 K^-$. $A(B^+ \to K^+ K^+ K^-)$ and $A(\bd
\to K^0 K^0 \kbar)$ vanish because there is no way of antisymmetrizing
the $K^+K^+$ or $K^0K^0$ pair. Here the final state has $I=\frac12$,
and there are two different ways of reaching it. We therefore expect
no relation between the amplitudes.

In order to write the amplitudes in terms of diagrams, we have to
antisymmetrize the $K^+$-$K^0$ state. As was done for $K\pi\pi$, we
adopt the following rule: all diagrams with the $K^+$-$K^0$ in order
of decreasing charge from top to bottom are unmodified; all diagrams
with the $K^+$-$K^0$ in order of increasing charge from top to bottom
get an additional factor of $-1$. The amplitudes (multiplied by
$\sqrt{2}$) are then given by
\bea
\label{KKKanti}
\sqrt{2} A(B^+ \to K^+ K^0 \kbar)_{anti} &=& -{\hat P}'_{uc} e^{i\gamma}+ {\hat P}'_{tc} \nn\\
&& \hskip0.8truecm -~\frac13 P'_{EW1,s} - \frac13 P'_{EW1} + \frac13
P^{\prime C}_{EW2,s} + \frac13 P^{\prime C}_{EW1} ~, \nn\\
\sqrt{2} A(\bd \to K^+ K^0 K^-)_{anti} &=& -T'_{2,s} e^{i\gamma}+C'_{1,s} e^{i\gamma}
-{\hat P}'_{uc} e^{i\gamma}+ {\hat P}'_{tc} \\
&& \hskip0.8truecm +~\frac23 P'_{EW1,s} - \frac13 P'_{EW1} - \frac23
P^{\prime C}_{EW2,s} + \frac13 P^{\prime C}_{EW1} ~. \nn
\eea
As expected, there is no relation between these two amplitudes.

\subsection{Penguin Dominance}

Assuming penguin dominance, Gronau and Rosner find that isospin
reflection implies the following equalities \cite{GR2005}:
\bea
A(B^+ \to K^+ K^+ K^-) &=& -A(\bd \to K^0 K^0 \kbar) ~, \nn\\
A(B^+ \to K^+ K^0 \kbar) &=& -A(\bd \to K^+ K^0 K^-) ~.
\eea
By distinguishing the symmetric and antisymmetric isospin states, it
is possible to go beyond these predictions. In the symmetric scenario,
if only ${\hat P}'_{tc}$ is retained, we predict
\bea
&& A(B^+ \to K^+ K^+ K^-) = -A(\bd \to K^0 K^0 \kbar) \nn\\
&& \hskip 2truecm =~- \sqrt{2} A(B^+ \to K^+ K^0 \kbar) = \sqrt{2} A(\bd \to K^+ K^0 K^-) ~.
\eea
(Note: the relations given in Eq.~(\ref{KKKapproxrels}) actually hold
for all diagrams, not just ${\hat P}'_{tc}$.) As discussed above, the
present data confirm the relation $A(B^+ \to K^+ K^+ K^-) = \sqrt{2}
A(\bd \to K^+ K^0 K^-)$.  In the antisymmetric scenario, we have only
$A(B^+ \to K^+ K^0 \kbar) = A(\bd \to K^+ K^0 K^-)$. As with $K\pi\pi$
decays, these provide further tests of the SM which.

\subsection{Isospin Amplitudes}

In Ref.~\cite{GR2003}, Gronau and Rosner (GR) write the amplitudes for
$B \to KK{\bar K}$ decays in terms of isospin amplitudes. It is
instructive to compare this with the diagrammatic description.

As described above, there are five independent isospin amplitudes,
denoted by $A_{\Delta I}^{I(KK),I_f} \equiv \bra{I(KK),I_f} \Delta I
\ket{\frac12}$, where $I(KK)$ is the isospin of the $KK$ pair [$I(KK)
  = 1$ (0) is symmetric (antisymmetric)], $I_f$ is the isospin of the
final state, and the weak Hamiltonian has $\Delta I = 0$ or 1. They
are listed as $A_0^{0,\frac12}$, $A_0^{1,\frac12}$, $A_1^{0,\frac12}$,
$A_1^{1,\frac12}$, $A_1^{1,\frac32}$.

As noted by GR, the $B \to KK{\bar K}$ amplitudes depend on the kaons'
momenta.  The amplitudes for $B^+ \to K^+ K^0 \kbar$ and $\bd \to K^+
K^0 K^-$ take different values when the $K^+$ and $K^0$ momenta are
exchanged. Thus, GR obtain expressions for six decay amplitudes in
terms of the five isospin amplitudes:
\bea
A(B^+ \to K^+ K^+ K^-)_{p_1 p_2 p_3} & = & 2 A_0^{1,\frac12} - 2 A_1^{1,\frac12} + A_1^{1,\frac32} ~, \nn\\
A(\bd \to K^0 K^0 \kbar)_{p_1 p_2 p_3} & = & - 2 A_0^{1,\frac12} - 2 A_1^{1,\frac12} + A_1^{1,\frac32} ~, \nn\\
A(B^+ \to K^+ K^0 \kbar)_{p_1 p_2 p_3} & = & A_0^{0,\frac12} - A_0^{1,\frac12} - A_1^{0,\frac12} + A_1^{1,\frac12} + A_1^{1,\frac32} ~, \nn\\
A(B^+ \to K^+ K^0 \kbar)_{p_2 p_1 p_3} & = & -A_0^{0,\frac12} - A_0^{1,\frac12} + A_1^{0,\frac12} + A_1^{1,\frac12} + A_1^{1,\frac32} ~, \nn\\
A(\bd \to K^+ K^0 K^-)_{p_1 p_2 p_3} & = & A_0^{0,\frac12} + A_0^{1,\frac12} + A_1^{0,\frac12} + A_1^{1,\frac12} + A_1^{1,\frac32} ~, \nn\\
A(\bd \to K^+ K^0 K^-)_{p_2 p_1 p_3} & = & - A_0^{0,\frac12} + A_0^{1,\frac12} - A_1^{0,\frac12} + A_1^{1,\frac12} + A_1^{1,\frac32} ~.
\eea
The above amplitudes are related to those of Eqs.~(\ref{KKKsym}) and
(\ref{KKKanti}) as follows:
\bea
A(B^+ \to K^+ K^+ K^-)_{sym} &=& A(B^+ \to K^+ K^+ K^-)_{p_1 p_2 p_3} ~, \nn\\
A(\bd \to K^0 K^0 \kbar)_{sym} &=& A(\bd \to K^0 K^0 \kbar)_{p_1 p_2 p_3} ~, \nn\\
\sqrt{2} A(B^+ \to K^+ K^0 \kbar)_{sym} &=& \nn\\
&& \hskip-1truein A(B^+ \to K^+ K^0 \kbar)_{p_1 p_2 p_3} + A(B^+ \to K^+ K^0 \kbar)_{p_2 p_1 p_3} ~, \nn\\
\sqrt{2} A(\bd \to K^+ K^0 K^-)_{sym} &=& \nn\\
&& \hskip-1truein A(\bd \to K^+ K^0 K^-)_{p_1 p_2 p_3} + A(\bd \to K^+ K^0 K^-)_{p_2 p_1 p_3} ~, \nn\\
\sqrt{2} A(B^+ \to K^+ K^0 \kbar)_{anti} &=& \nn\\
&& \hskip-1truein A(B^+ \to K^+ K^0 \kbar)_{p_1 p_2 p_3} - A(B^+ \to K^+ K^0 \kbar)_{p_2 p_1 p_3} ~, \nn\\
\sqrt{2} A(\bd \to K^+ K^0 K^-)_{anti} &=& \nn\\
&& \hskip-1truein A(\bd \to K^+ K^0 K^-)_{p_1 p_2 p_3} - A(\bd \to K^+ K^0 K^-)_{p_2 p_1 p_3} ~.
\eea

Now, because there are six decay amplitudes, but only five isospin
amplitudes, there must be a relation between the decay amplitudes. GR
give this relation as
\bea
&& A(B^+ \to K^+ K^+ K^-)_{p_1 p_2 p_3} + A(B^+ \to K^+ K^0 \kbar)_{p_1 p_2 p_3} \nn\\
&& \hskip2.3truein +~A(B^+ \to K^+ K^0 \kbar)_{p_2 p_1 p_3} = \nn\\
&& A(\bd \to K^0 K^0 \kbar)_{p_1 p_2 p_3} + A(\bd \to K^+ K^0 K^-)_{p_1 p_2 p_3} \nn\\
&& \hskip2.3truein +~A(\bd \to K^+ K^0 K^-)_{p_2 p_1 p_3} = 3A_1^{1,\frac32} ~.
\eea
This is the same as the relation in Eq.~(\ref{KKKrel}). However, when
one expresses the amplitudes in terms of diagrams, there are, in fact,
two relations instead of one [Eq.~(\ref{KKKapproxrels})]. This implies
that
\beq
A_1^{1,\frac12} =-\frac14 A_1^{1,\frac32} ~,
\eeq
so that there are really four independent isospin amplitudes instead
of five.  As described above, the extra relation is a consequence of
neglecting the annihilation-type diagrams. In other words, the above
relation among isospin amplitudes is a good approximation, and could
not have been deduced without performing a diagrammatic analysis.

It is straightforward to express the remaining isospin amplitudes in
terms of diagrams:
\bea
A_0^{1,\frac12} &=& \frac14 \left[ -T'_{2,s} e^{i\gamma}-C'_{1,s} e^{i\gamma}
-2{\hat P}'_{uc} e^{i\gamma}+ 2{\hat P}'_{tc} \right. \nn\\
&& \hskip0.8truecm \left. +~\frac13 P'_{EW1,s} - \frac23 P'_{EW1} + \frac13
P^{\prime C}_{EW2,s} - \frac23 P^{\prime C}_{EW1} \right] ~, \nn\\
A_1^{1,\frac32} &=& \frac13 \left[ -T'_{2,s} e^{i\gamma}-C'_{1,s} e^{i\gamma} +  P'_{EW1,s} + P^{\prime C}_{EW2,s} \right] ~, \nn\\
A_0^{0,\frac12} &=& \frac14 \left[ -T'_{2,s} e^{i\gamma}+C'_{1,s} e^{i\gamma} -2{\hat P}'_{uc} e^{i\gamma}+ 2{\hat P}'_{tc} \right. \nn\\
&& \hskip0.8truecm \left. +~\frac13 P'_{EW1,s} - \frac23 P'_{EW1} - \frac13
P^{\prime C}_{EW2,s} + \frac23 P^{\prime C}_{EW1} \right] ~, \nn\\
A_1^{0,\frac12} &=& \frac14 \left[ -T'_{2,s} e^{i\gamma}+C'_{1,s} e^{i\gamma} + P'_{EW1,s} - P^{\prime C}_{EW2,s} \right] ~,
\eea
(Recall that, despite their having the same name, the diagrams which
contribute to the $A_{\{0,1\}}^{1,\{\frac12.\frac32\}}$ and
$A_{\{0,1\}}^{0,\frac12}$ isospin amplitudes are not the same -- they
can have different sizes.) In the limit of penguin dominance,
$A_1^{1,\frac32}$ and $A_0^{1,\frac12}$ vanish. This is consistent
with what is found in the previous subsection.

\subsection{Weak-Phase Information}
\label{KKKweak}

As was the case for $B \to K\pi\pi$ decays, the amplitudes contain the
weak phase $\gamma$, and so one wonders if it can be measured in $B
\to KK{\bar K}$ decays. Here the answer is `perhaps'.

When the isospin state of the $KK$ pair is symmetric, there are four
decays. However, due to the equality relations in
Eq.~(\ref{KKKapproxrels}), two of these have the same amplitudes as
the other two. There are therefore 6 observables: the
momentum-dependent branching ratios, direct CP asymmetries and
indirect CP asymmetries of of $\bd \to K^+ K^0 K^-$ and $\bd \to K^0
K^0 \kbar$. In the antisymmetric scenario, there are 5 observables:
the momentum-dependent branching ratios and direct CP asymmetries of
$B^+ \to K^+ K^0 \kbar$ and $\bd \to K^+ K^0 K^-$, and the
momentum-dependent indirect CP asymmetry of $\bd \to K^+ K^0 K^-$.
(As with $B \to K\pi\pi$, the separation of symmetric and
antisymmetric $KK$ states fixes the CP of the final state for the
indirect CP asymmetries.)

However, in either case, the amplitudes [Eqs.~(\ref{KKKsym}) and
  (\ref{KKKanti})] are written in terms of 4 effective diagrams,
corresponding to 8 theoretical parameters: 4 magnitudes of diagrams, 3
relative (strong) phases, and $\gamma$. This is larger than the number
of observables, and so the weak phase $\gamma$ cannot be extracted
from $B \to KK{\bar K}$ decays.

The best that one can do is to assume the hierarchy of
Eq.~(\ref{hierarchy}), and neglect all $C'$, ${\hat P}'_{uc}$ and
$P^{\prime C}_{EW}$ diagrams. This reduces the number of effective
diagrams to 3, which corresponds to 6 theoretical parameters. This is
equal to the number of observables in the symmetric case, so that
$\gamma$ can be extracted here, albeit with discrete ambiguities. And,
as described above, the theoretical error is uncertain.

\section{\boldmath $B \to K{\bar K}\pi$ Decays}

We now consider $B \to K{\bar K}\pi$ decays, which are $\btod$
transitions. Here there are seven processes: $B^+ \to K^+K^-\pi^+$,
$B^+ \to K^+\kbar\pi^0$, $B^+ \to K^0\kbar\pi^+$, $\bd \to
K^+K^-\pi^0$, $\bd \to K^+\kbar\pi^-$, $\bd \to K^0\kbar\pi^0$, $\bd
\to K^0K^-\pi^+$. There are no identical particles in the final state,
so here we do not have to distinguish symmetric and antisymmetric
isospin states.

In $B \to K{\bar K}\pi$, the final state has $I=0$, $I=1$ (twice) or
$I=2$.  The weak Hamiltonian has $\Delta I = \frac12$ or $\frac32$, so
there are six paths to the final state. This implies that there is one
relation among the seven decay amplitudes. It is
\bea
&& \sqrt{2} A(\bd \to K^+K^-\pi^0) 
+ A(\bd\to K^0K^-\pi^+)
- A(B^+ \to K^+K^-\pi^+) \nn\\
&& \hskip 2truecm +~\sqrt{2} A(\bd \to K^0\kbar\pi^0)
+ A(\bd \to K^+\kbar\pi^-) \nn\\
&& \hskip 2truecm -~A(B^+ \to K^0\kbar\pi^+)
- \sqrt{2} A(B^+ \to K^+\kbar\pi^0) = 0 ~.
\label{KKpirel}
\eea

In terms of diagrams, the amplitudes are given by
\bea
\label{KKpiamps}
A(B^+ \to K^+K^-\pi^+) &=& \left[ T_{2,s} + C_{1,s}
+ P_{a;uc} \right] e^{-i\alpha} \nn\\ 
&& \hskip-1truecm -~P_{a;tc}  + \frac13 P_{EW1}  -\frac23 P_{EW1,s}  + \frac13
P^C_{EW1}  - \frac23 P^C_{EW2,s} ~, \nn\\
\sqrt{2} A(B^+ \to K^+\kbar\pi^0) &=& \left[ T_{1,s} + C_{2,s}
- P_{a;uc}  + P_{b;uc}  \right] e^{-i\alpha} \nn\\
&& \hskip-4truecm +~P_{a;tc}  - P_{b;tc}  -~ P_{EW2,s}  - \frac13 P^C_{EW1}  - \frac23
P^C_{EW1,s}  + \frac13 P^C_{EW2}  - \frac13 P^C_{EW2,s}  ~, \nn\\
A(B^+ \to K^0\kbar\pi^+) &=& -P_{b;uc} e^{-i\alpha} \nn\\
&& \hskip-1truecm +~P_{b;tc}  -~\frac13 P_{EW1}  -\frac13 P_{EW1,s}  - \frac13
P^C_{EW1,s}  - \frac13 P^C_{EW2} ~, \nn\\
\sqrt{2} A(\bd \to K^+K^-\pi^0) &=& C_{1,s} e^{-i\alpha} + \frac13
P_{EW1} -\frac23 P_{EW1,s} ~, \nn\\
A(\bd \to K^+\kbar\pi^-) &=& \left[ T_{1,s} + P_{b;uc} \right]
e^{-i\alpha}  - P_{b;tc} -~\frac23 P^C_{EW1,s} + \frac13 P^C_{EW2} ~, \nn\\
\sqrt{2} A(\bd \to K^0\kbar\pi^0) &=& \left[ C_{2,s} - P_{a;uc} 
-P_{b;uc} \right] e^{-i\alpha} \\
&& \hskip0.2truecm +~P_{a;tc}  + P_{b;tc}  -~\frac13 P_{EW1} -\frac13 P_{EW1,s}  -P_{EW2,s}  
\nn\\
&& \hskip0.5truecm -~
\frac13 P^C_{EW1}  -~\frac13 P^C_{EW1,s}  - \frac13 P^C_{EW2}  - \frac13
P^C_{EW2,s}  ~, \nn\\
A(\bd\to K^0K^-\pi^+) &=& \left[ T_{2,s}  + P_{a;uc} \right]
e^{-i\alpha} - P_{a;tc}  +~\frac13 P^C_{EW1} - \frac23 P^C_{EW2,s} ~, \nn
\eea
where $P_a \equiv P_1 + P_{2,s}$, $P_b \equiv P_{1,s} + P_2$, and all
amplitudes have been multiplied by $e^{i\beta}$. With these
expressions, the relation of Eq.~(\ref{KKpirel}) is reproduced.

However, there are, in fact, two relations:
\bea
\sqrt{2} A(\bd \to K^+K^-\pi^0) 
+ A(\bd\to K^0K^-\pi^+)
&=&  A(B^+ \to K^+K^-\pi^+) ~, \nn\\
&& \hskip -8truecm \sqrt{2} A(\bd \to K^0\kbar\pi^0)
+ A(\bd \to K^+\kbar\pi^-) \nn\\
&& \hskip -6.5truecm =~ A(B^+ \to K^0\kbar\pi^+)
+ \sqrt{2} A(B^+ \to K^+\kbar\pi^0) ~.
\eea
As was the case in $B \to KK{\bar K}$ decays, the (justified) neglect
of certain annihilation-type diagrams breaks the relation in
Eq.~(\ref{KKpirel}) into two.

\subsection{\boldmath $T$ Dominance}

In two-body $B$ decays, $T$ is the dominant diagram in $\btod$
transitions. Assuming this also holds in three-body $B$ decays, we
have the following predictions:
\bea
A(B^+ \to K^+K^-\pi^+) &=& A(\bd\to K^0K^-\pi^+) ~, \nn\\
\sqrt{2} A(B^+ \to K^+\kbar\pi^0) &=& A(\bd \to K^+\kbar\pi^-) ~,
\nn\\
&& \hskip-6.5truecm A(B^+ \to K^0\kbar\pi^+) = A(\bd \to K^+K^-\pi^0) =
A(\bd \to K^0\kbar\pi^0) \simeq 0 ~.
\eea
These are tests of the SM which can be carried out once these decays
are measured.

\subsection{Weak-Phase Information}
\label{KKpialpha}

There are seven $B \to K{\bar K}\pi$ decays, which yield 16
observables: the branching ratios and direct CP asymmetries of $B^+
\to K^+K^-\pi^+$, $B^+ \to K^+\kbar\pi^0$, $B^+ \to K^0\kbar\pi^+$,
$\bd \to K^+K^-\pi^0$, $\bd \to K^+\kbar\pi^-$, $\bd \to
K^0\kbar\pi^0$, $\bd \to K^0K^-\pi^+$, and the indirect CP asymmetries
of $\bd \to K^+K^-\pi^0$, $\bd \to K^0\kbar\pi^0$.

The $B \to K{\bar K}\pi$ amplitudes in Eq.~(\ref{KKpiamps}) can be
written in terms of 10 effective diagrams:
\bea
A(B^+ \to K^+K^-\pi^+) &=& [D_1 + D_3] e^{-i\alpha} + D_2 + D_4  ~, \nn\\
\sqrt{2} A(B^+ \to K^+\kbar\pi^0) &=& D_9 e^{-i\alpha} + D_{10}  ~, \nn\\
A(B^+ \to K^0\kbar\pi^+) &=& D_7 e^{-i\alpha} + D_8  ~, \nn\\
\sqrt{2} A(\bd \to K^+K^-\pi^0) &=& D_1 e^{-i\alpha} + D_2  ~, \nn\\
A(\bd \to K^+\kbar\pi^-) &=&  D_5 e^{-i\alpha} + D_6  ~, \nn\\
\sqrt{2} A(\bd \to K^0\kbar\pi^0) &=& [-D_5 + D_7 + D_9] e^{-i\alpha} - D_6 + D_8 + D_{10}   ~, \nn\\
A(\bd\to K^0K^-\pi^+) &=&  D_3 e^{-i\alpha} + D_4  ~,
\eea
where
\bea
D_1 &\equiv& C_{1,s} ~, \nn\\
D_2 &\equiv& \frac13 P_{EW1} -\frac23 P_{EW1,s} ~, \nn\\
D_3 &\equiv& T_{2,s} + P_{a;uc} ~, \nn\\
D_4 &\equiv& - P_{a;tc} + \frac13 P^C_{EW1} -\frac23 P^C_{EW2,s} ~, \nn\\
D_5 &\equiv& T_{1,s} + P_{b;uc} ~, \nn\\
D_6 &\equiv& - P_{b;tc} + \frac13 P^C_{EW2} - \frac23 P^C_{EW1,s}  ~, \nn\\
D_7 &\equiv& -P_{b;uc} ~, \nn\\
D_8 &\equiv& P_{b;tc} - \frac13 P_{EW1} -\frac13 P_{EW1,s} - \frac13 P^C_{EW2} -\frac13 P^C_{EW1,s} ~, \nn\\
D_9 &\equiv& T_{1,s} + C_{2,s} - P_{a;uc} + P_{b;uc} ~, \nn\\
D_{10} &\equiv& P_{a;tc} - P_{b;tc} -P_{EW2,s} - \frac13 P^C_{EW1} - \frac23 P^C_{EW1,s} + \frac13 P^C_{EW2} - \frac13 P^C_{EW2,s} ~.
\eea
This corresponds to 20 theoretical parameters: 10 magnitudes of
diagrams, 9 relative (strong) phases, and $\alpha$. With only 16
observables, $\alpha$ cannot be extracted.

We therefore need additional input. Fortunately, we have some, similar
to that in Secs.~\ref{Kpipiweak} and \ref{KKKweak}. In two-body
$\btod$ $B$ decays, the diagrams obey the approximate hierarchy
\cite{GHLR}
\bea
\label{btodhierarchy}
1 &:& T ~, \nn\\
{\bar\lambda} &:& C, P_{tc}, P_{uc} ~, \nn\\
{\bar\lambda}^2 &:& P_{EW} ~, \nn\\
{\bar\lambda}^3 &:& P^C_{EW} ~.
\eea
If the three-body decay diagrams obey a similar hierarchy, all EWP
diagrams can be neglected, leading to an error of only $\sim 5\%$. In
this limit, we have $D_2 = 0$, $D_8 = -D_6$, and $D_{10} = -D_4 +
D_6$. So the number of independent diagrams is reduced to 7, i.e.\ 14
theoretical parameters\footnote{We assume that, for the indirect CP
  asymmetries, the CP of the final state can be fixed as for the
  decays in previous sections. Otherwise there are 2 additional
  theoretical parameters.}. Thus, by measuring the observables in $B
\to K{\bar K}\pi$ decays, weak-phase information can be obtained. In
fact, not all 16 observables are necessary.  Experimentally, this is
not easy, but it is at least theoretically possible. Of course, as in
Secs.~\ref{Kpipiweak} and \ref{KKKweak}, the theoretical error is
uncertain, since it is difficult to test the hierarchy of diagrams.

\section{\boldmath $B \to \pi\pi\pi$ Decays}

Finally, we examine $B \to \pi\pi\pi$ decays, also a $\btod$
transition. There are four processes: $\bd \to \pi^0\pi^0\pi^0$, $B^+
\to \pi^+\pi^0\pi^0$, $B^+ \to \pi^-\pi^+\pi^+$, $\bd \to
\pi^+\pi^0\pi^-$. In contrast to the other decays, here the final
state includes three identical particles under isospin, so that the
six permutations of these particles (the group $S_3$) must be
considered.  Numbering the particles 1, 2, 3, the six possible orders
are 123, 132, 312, 321, 231, 213. Under $S_3$, there are six
possibilities for the isospin state of the three $\pi$'s: a totally
symmetric state $\ket{S}$, a totally antisymmetric state $\ket{A}$, or
one of four mixed states $\ket{M_i}$ ($i=1$-4). These can be defined
as
\bea
\ket{S} &\equiv& \frac{1}{\sqrt{6}} \left( \ket{123} + \ket{132} + \ket{312} + \ket{321} + \ket{231} + \ket{213} \right)~,\nn\\
\ket{M_1} &\equiv& \frac{1}{\sqrt{12}} \left( 2\ket{123} + 2\ket{132} - \ket{312} - \ket{321} - \ket{231} - \ket{213} \right)~,\nn\\
\ket{M_2} &\equiv& \frac{1}{\sqrt{4}} \left( \ket{312} - \ket{321} - \ket{231} + \ket{213} \right)~,\nn\\
\ket{M_3} &\equiv& \frac{1}{\sqrt{4}} \left( -\ket{312} - \ket{321} + \ket{231} + \ket{213} \right)~,\nn\\
\ket{M_4} &\equiv& \frac{1}{\sqrt{12}} \left( 2\ket{123} - 2\ket{132} - \ket{312} + \ket{321} - \ket{231} + \ket{213} \right)~,\nn\\
\ket{A} &\equiv& \frac{1}{\sqrt{6}} \left( \ket{123} - \ket{132} + \ket{312} - \ket{321} + \ket{231} - \ket{213} \right)~.
\label{SU3states}
\eea
This choice of mixed states implies that two truly identical particles
go in positions 2 and 3. Under the exchange $2\leftrightarrow 3$,
$\ket{M_1}$ and $\ket{M_2}$ are symmetric, while $\ket{M_3}$ and
$\ket{M_4}$ are antisymmetric.

For the four $B\to\pi\pi\pi$ decays, we have:
\begin{enumerate}

\item $\bd\to \pi^0\pi^0\pi^0$: all final-state particles are the
  same, which means $\ket{123} = \ket{132} = \ket{312} = \ket{321} =
  \ket{231} = \ket{213}$. In this case, only the state $\ket{S}$ is
  allowed.

\item $B^+\to \pi^+\pi^0\pi^0$: particle 1 is $\pi^+$, particles 2 and
  3 are $\pi^0$. Thus, $\ket{123} = \ket{132}$, $\ket{312} =
  \ket{213}$, $\ket{231} = \ket{321}$. This implies that each of
  $\ket{M_3}$, $\ket{M_4}$, $\ket{A}$ is not allowed.

\item $B^+\to \pi^-\pi^+\pi^+$: particle 1 is $\pi^-$, particles 2 and
  3 are $\pi^+$. Thus, $\ket{123} = \ket{132}$, $\ket{312} =
  \ket{213}$, $\ket{231} = \ket{321}$. This implies that each of
  $\ket{M_3}$, $\ket{M_4}$, $\ket{A}$ is not allowed.

\item $\bd\to \pi^+\pi^0\pi^-$: we choose the order such that particle
  1 is $\pi^+$, particle 2 is $\pi^0$, particle 3 is $\pi^-$. All six
  states are allowed.

\end{enumerate}
The amplitude for a decay with two truly identical particles has an
extra factor of $1/\sqrt{2}$; with three truly identical particles,
the factor is $1/\sqrt{6}$.

The six elements of $S_3$ are: $I$ (identity), $P_{12}$ (exchanges
particles 1 and 2), $P_{13}$ (exchanges particles 1 and 3), $P_{23}$
(exchanges particles 2 and 3), $P_{cyclic}$ (cyclic permutation of
particle numbers, i.e.\ $1\to 2$, $2\to 3$, $3\to 1$),
$P_{anticyclic}$ (anticyclic permutation of particle numbers,
i.e.\ $1\to 3$, $2\to 1$, $3\to 2$). Under the group transformations,
$\ket{S} \to \ket{S}$ and $\ket{A} \to \pm\ket{A}$. It is easy to see
that $\ket{M_1}$ and $\ket{M_3}$ transform among themselves. Writing
\beq
\ket{M_1} \equiv \left( \matrix{1 \cr 0} \right) ~~,~~~~ 
\ket{M_3} \equiv \left( \matrix{0 \cr 1} \right) ~~,
\eeq
we can represent each group element by a $2\times 2$ matrix:
\bea
& I = \left( \matrix{1 & 0 \cr 0 & 1} \right) ~,~~
P_{12} = \left( \matrix{-\frac12 & \frac{\sqrt{3}}{2} \cr \frac{\sqrt{3}}{2} & \frac12} \right) ~,~~
P_{13} = \left( \matrix{-\frac12 & -\frac{\sqrt{3}}{2} \cr -\frac{\sqrt{3}}{2} & \frac12} \right) ~, & \nn\\
& P_{23} = \left( \matrix{1 & 0 \cr 0 & -1} \right) ~,~~
P_{cyclic} = \left( \matrix{-\frac12 & \frac{\sqrt{3}}{2} \cr -\frac{\sqrt{3}}{2} & -\frac12} \right) ~,~~
P_{anticyclic} = \left( \matrix{-\frac12 & -\frac{\sqrt{3}}{2} \cr \frac{\sqrt{3}}{2} & -\frac12} \right) ~. &
\label{matrices}
\eea
Similarly, if we write
\beq
\ket{M_2} \equiv \left( \matrix{1 \cr 0} \right) ~~,~~~~ 
\ket{M_4} \equiv \left( \matrix{0 \cr 1} \right) ~~,
\eeq
the $S_3$ matrices take the same form, showing that $\ket{M_2}$ and
$\ket{M_4}$ also transform among themselves.

The above allows us to express the amplitudes for all $B\to\pi\pi\pi$
decays in terms of diagrams.  We begin with some general comments
about diagrams. As an example, consider $T_1$. In principle, there are
six possibilities, $T_1^{ijk}$, in which the final-state pions $i$,
$j$, $k$ run from top to bottom of the diagram in all
permutations. Suppose that we want the expression for the amplitude of
$B\to\pi_1\pi_2\pi_3$ in a particular $\ket{S_3}$ state, and suppose
that the diagram $T_1^{ijk}$ contributes to the decay. For $\ket{S_3}
= \ket{S}$, we define $T_1^S$:
\beq
T_1^S \equiv \frac{1}{\sqrt{6}} \left( T_1^{123} + T_1^{132} +
T_1^{312} + T_1^{321} + T_1^{231} + T_1^{213} \right) ~.
\eeq
Each $T_1^{ijk}$ leads to $T_1^S$ in the amplitude. For
$\ket{S_3} = \ket{A}$, we have
\beq
T_1^A \equiv \frac{1}{\sqrt{6}} \left( T_1^{123} - T_1^{132} +
T_1^{312} - T_1^{321} + T_1^{231} - T_1^{213} \right) ~.
\eeq
Again, each $T_1^{ijk}$ leads to $T_1^A$ in the amplitude, with a
coefficient of 1 ($-1$) if $ijk$ is in cyclic (anticyclic) order.

For the mixed states, one has to take into account the fact that,
under group transformations, there is $\ket{M_1}$-$\ket{M_3}$ and
$\ket{M_2}$-$\ket{M_4}$ mixing. In order to illustrate how this is
done, we focus first on the $M_1$/$M_3$ sector. We define
\bea
T_1^{M_1} &\equiv& \frac{1}{\sqrt{12}} \left( 2T_1^{123} + 2T_1^{132}
- T_1^{312} - T_1^{321} - T_1^{231} - T_1^{213} \right)~,\nn\\
T_1^{M_3} &\equiv& \frac{1}{\sqrt{4}} \left( - T_1^{312} - T_1^{321} +
T_1^{231} + T_1^{213} \right)~.
\eea
Suppose $\ket{S_3} = \ket{M_1}$.  The contribution to the amplitude of
$B\to\pi_1\pi_2\pi_3$ is $[M \times
  (T_1^{M_1},T_1^{M_3})^T]_{upper~component}$, where $M$ is the matrix
representing the $S_3$ group element which transforms $ijk$ to 123
[Eq.~(\ref{matrices})]. In general, this is a combination of
$T_1^{M_1}$ and $T_1^{M_3}$ (though the $T_1^{M_3}$ component can be
zero if $M=I$ or $P_{23}$). Factors of $-1$ for each ${\bar u}$ and
$1/\sqrt{2}$ for each $\pi^0$ must also be included. If $\ket{S_3} =
\ket{M_3}$, the contribution to the amplitude is $[M \times
  (T_1^{M_1},T_1^{M_3})^T]_{lower~component}$.  This can be applied
analogously to the $M_2$/$M_4$ sector, where we define
\bea
T_1^{M_2} &\equiv& \frac{1}{\sqrt{4}} \left( T_1^{312} - T_1^{321} -
T_1^{231} + T_1^{213} \right)~, \nn\\
T_1^{M_4} &\equiv& \frac{1}{\sqrt{12}} \left( 2T_1^{123} - 2T_1^{132}
- T_1^{312} + T_1^{321} - T_1^{231} + T_1^{213} \right)~.
\eea
The entire procedure holds for all diagrams\footnote{When applied to
  the decays in the previous sections, this method produces the same
  amplitude decomposition as when we used the simple rule of adding a
  minus sign to diagrams in which the identical particles are
  exchanged (e.g. in $B\to K\pi\pi$ or $KK{\bar K}$).}.

With these rules, we can now work out the amplitudes for all
decays. We begin first with $\ket{S_3} = \ket{S}$. The amplitudes are
\bea
\frac{2}{\sqrt{3}} A(\bd\to \pi^0\pi^0\pi^0)_{\ket{S}} &=&
- \left[ C_1^S - C_2^S + P^S_{uc} \right] e^{-i\alpha} \nn\\
&& \hskip-1truecm +~\left[ P^S_{tc} +~\frac13 P_{EW1}^S - P_{EW2}^S - \frac13 P^{C,S}_{EW1} - \frac13
P^{C,S}_{EW2} \right]  ~, \nn\\
\sqrt{2} A(B^+\to \pi^+\pi^0\pi^0)_{\ket{S}} &=& - \left[ T_2^S + C_1^S + P^S_{uc} \right] e^{-i\alpha} \nn\\
&& \hskip-1truecm +~\left[ P^S_{tc} +~\frac13 P_{EW1}^S -\frac13 P^{C,S}_{EW1} + \frac23 P^{C,S}_{EW2} \right]  ~, \nn\\
\frac{1}{\sqrt{2}} A(B^+\to \pi^-\pi^+\pi^+)_{\ket{S}} &=& \left[ T_2^S + C_1^S + P^S_{uc} \right] e^{-i\alpha} \nn\\
&& \hskip-1truecm -~\left[ P^S_{tc} + \frac13 P_{EW1}^S - \frac13 P^{C,S}_{EW1} + \frac23 P^{C,S}_{EW2} \right]  ~, \nn\\
\sqrt{2} A(\bd\to \pi^+\pi^0\pi^-)_{\ket{S}} &=& \left[ C_1^S - C_2^S + P^S_{uc} \right] e^{-i\alpha} \nn\\
&& \hskip-1truecm -~\left[ P^S_{tc} + \frac13 P_{EW1}^S - P_{EW2}^S - \frac13 P^{C,S}_{EW1} - \frac13 P^{C,S}_{EW2} \right]  ~,
\eea
where $P \equiv P_1 +P_2$ and all amplitudes have been multiplied by
$e^{i\beta}$.

For the $M_1$/$M_3$ sector, the amplitudes are
\bea
\sqrt{2} A(B^+\to \pi^+\pi^0\pi^0)_{\ket{M_1}} &=& \left[ \frac32
  T_1^{M_1} - \frac{\sqrt{3}}{2} T_1^{M_3} - T_2^{M_1} - C_1^{M_1} +
  \frac32 C_2^{M_1} - \frac{\sqrt{3}}{2} C_2^{M_3} \right. \nn\\
&& \hskip-1.5truein \left.  -~P^{M_1}_{uc} + \sqrt{3} P^{M_3}_{uc}
  \right] e^{-i\alpha} + \left[ P^{M_1}_{tc} - \sqrt{3} P^{M_3}_{tc} -
  \frac16 P_{EW1}^{M_1} - \frac{1}{2\sqrt{3}} P_{EW1}^{M_3} \right. \nn\\
&& \hskip-1truein \left. +~\sqrt{3} P_{EW2}^{M_3} - \frac13
  P^{C,M_1}_{EW1} - \frac{2}{\sqrt{3}} P^{C,M_3}_{EW1} - \frac56
  P^{C,M_1}_{EW2} - \frac{1}{2\sqrt{3}} P^{C,M_3}_{EW2} \right] ~,
\nn\\
\sqrt{2} A(B^+\to \pi^-\pi^+\pi^+)_{\ket{M_1}} &=& \left[ - T_2^{M_1} +
\sqrt{3} T_2^{M_3} - C_1^{M_1} - \sqrt{3} C_1^{M_3} \right. \nn\\
&& \hskip-1.5truein \left. -~P^{M_1}_{uc} + \sqrt{3} P^{M_3}_{uc} \right] e^{-i\alpha}
+ \left[ P^{M_1}_{tc} - \sqrt{3} P^{M_3}_{tc} + \frac43 P_{EW1}^{M_1} -
\frac{2}{\sqrt{3}} P_{EW1}^{M_3} \right. \nn\\
&& \hskip-1truein \left. -~\frac13 P^{C,M_1}_{EW1} + \frac{1}{\sqrt{3}}
P^{C,M_3}_{EW1} + \frac23 P^{C,M_1}_{EW2} - \frac{2}{\sqrt{3}}
P^{C,M_3}_{EW2} \right] ~, \nn\\
6 \sqrt{2} A(\bd\to \pi^+\pi^0\pi^-)_{\ket{M_1}} &=& \left[ 9
  T_1^{M_1} - 3\sqrt{3} T_1^{M_3} - 3C_1^{M_1} + 3\sqrt{3} C_1^{M_3} +
  3C_2^{M_1} \right. \nn\\
&& \hskip-1.5truein \left. -~3\sqrt{3} C_2^{M_3} - 3 P^{M_1}_{uc} + 3 \sqrt{3} P^{M_3}_{uc}
  \right] e^{-i\alpha} + \left[ 3 P^{M_1}_{tc} - 3 \sqrt{3} P^{M_3}_{tc} \right. \nn\\
&& \hskip-1truein -~5 P_{EW1}^{M_1} + \sqrt{3} P_{EW1}^{M_3} -3 P_{EW2}^{M_1} + 3 \sqrt{3} P_{EW2}^{M_3} \nn\\
&& \hskip-1truein 
\left. -~P^{C,M_1}_{EW1} - 5\sqrt{3} P^{C,M_3}_{EW1} -
  P^{C,M_1}_{EW2} + \sqrt{3} P^{C,M_3}_{EW2} \right]
~, \nn\\
2 \sqrt{6} A(\bd\to \pi^+\pi^0\pi^-)_{\ket{M_3}} &=& \left[ - 3 T_1^{M_1} +
\sqrt{3} T_1^{M_3} - 4\sqrt{3} T_2^{M_3} + 3 C_1^{M_1} + \sqrt{3}
C_1^{M_3} \right. \nn\\
&& \hskip-1.5truein \left. +~3 C_2^{M_1} + \sqrt{3} C_2^{M_3} + 3
P^{M_1}_{uc} - 3\sqrt{3} P^{M_3}_{uc} \right] e^{-i\alpha} + \left[ - 3 P^{M_1}_{tc} + 3\sqrt{3}
P^{M_3}_{tc} \right. \nn\\
&& \hskip-1truein -~P_{EW1}^{M_1} + \sqrt{3} P_{EW1}^{M_3} +~3
P_{EW2}^{M_1} + \sqrt{3} P_{EW2}^{M_3} \nn\\
&& \hskip-1truein
\left. +~P^{C,M_1}_{EW1} + \sqrt{3} P^{C,M_3}_{EW1} - 5
P^{C,M_1}_{EW2} + \sqrt{3} P^{C,M_3}_{EW2} \right] ~.
\eea

For the $M_2$/$M_4$ sector, the amplitudes are
\bea
\sqrt{2} A(B^+\to \pi^+\pi^0\pi^0)_{\ket{M_2}} &=& \left[ \frac32
  T_1^{M_2} - \frac{\sqrt{3}}{2} T_1^{M_4} - T_2^{M_2} - C_1^{M_2} +
  \frac32 C_2^{M_2} - \frac{\sqrt{3}}{2} C_2^{M_4} \right. \nn\\
&& \hskip-1.5truein \left.  -~P^{M_2}_{uc} + \sqrt{3} P^{M_4}_{uc}
  \right] e^{-i\alpha} + \left[ P^{M_2}_{tc} - \sqrt{3} P^{M_4}_{tc} -
  \frac16 P_{EW1}^{M_2} - \frac{1}{2\sqrt{3}} P_{EW1}^{M_4} \right. \nn\\
&& \hskip-1truein \left. +~\sqrt{3} P_{EW2}^{M_4} - \frac13
  P^{C,M_2}_{EW1} - \frac{2}{\sqrt{3}} P^{C,M_4}_{EW1} - \frac56
  P^{C,M_2}_{EW2} - \frac{1}{2\sqrt{3}} P^{C,M_4}_{EW2} \right] ~,
\nn\\
\sqrt{2} A(B^+\to \pi^-\pi^+\pi^+)_{\ket{M_2}} &=& \left[ - T_2^{M_2} +
\sqrt{3} T_2^{M_4} - C_1^{M_2} - \sqrt{3} C_1^{M_4} \right. \nn\\
&& \hskip-1.5truein \left. -~P^{M_2}_{uc} + \sqrt{3} P^{M_4}_{uc} \right] e^{-i\alpha}
+ \left[ P^{M_2}_{tc} - \sqrt{3} P^{M_4}_{tc} + \frac43 P_{EW1}^{M_2} -
\frac{2}{\sqrt{3}} P_{EW1}^{M_4} \right. \nn\\
&& \hskip-1truein \left. -~\frac13 P^{C,M_2}_{EW1} + \frac{1}{\sqrt{3}}
P^{C,M_4}_{EW1} + \frac23 P^{C,M_2}_{EW2} - \frac{2}{\sqrt{3}}
P^{C,M_4}_{EW2} \right] ~, \nn\\
6 \sqrt{2} A(\bd\to \pi^+\pi^0\pi^-)_{\ket{M_2}} &=& \left[ 9
  T_1^{M_2} - 3\sqrt{3} T_1^{M_4} - 3C_1^{M_2} + 3\sqrt{3} C_1^{M_4} +
  3C_2^{M_2} \right. \nn\\
&& \hskip-1.5truein \left. -~3\sqrt{3} C_2^{M_4} - 3 P^{M_2}_{uc} + 3 \sqrt{3} P^{M_4}_{uc}
  \right] e^{-i\alpha} + \left[ 3 P^{M_2}_{tc} - 3 \sqrt{3} P^{M_4}_{tc} \right. \nn\\
&& \hskip-1truein -~5 P_{EW1}^{M_2} + \sqrt{3} P_{EW1}^{M_4} -3 P_{EW2}^{M_2} + 3 \sqrt{3} P_{EW2}^{M_4} \nn\\
&& \hskip-1truein 
\left. -~P^{C,M_2}_{EW1} - 5\sqrt{3} P^{C,M_4}_{EW1} -
  P^{C,M_2}_{EW2} + \sqrt{3} P^{C,M_4}_{EW2} \right]
~, \nn\\
2 \sqrt{6} A(\bd\to \pi^+\pi^0\pi^-)_{\ket{M_4}} &=& \left[ - 3 T_1^{M_2} +
\sqrt{3} T_1^{M_4} - 4\sqrt{3} T_2^{M_4} + 3 C_1^{M_2} + \sqrt{3}
C_1^{M_4} \right. \nn\\
&& \hskip-1.5truein \left. +~3 C_2^{M_2} + \sqrt{3} C_2^{M_4} + 3
P^{M_2}_{uc} - 3\sqrt{3} P^{M_4}_{uc} \right] e^{-i\alpha} + \left[ - 3 P^{M_2}_{tc} + 3\sqrt{3}
P^{M_4}_{tc} \right. \nn\\
&& \hskip-1truein -~P_{EW1}^{M_2} + \sqrt{3} P_{EW1}^{M_4} +~3
P_{EW2}^{M_2} + \sqrt{3} P_{EW2}^{M_4} \nn\\
&& \hskip-1truein
\left. +~P^{C,M_2}_{EW1} + \sqrt{3} P^{C,M_4}_{EW1} - 5
P^{C,M_2}_{EW2} + \sqrt{3} P^{C,M_4}_{EW2} \right] ~.
\eea

Finally, for $\ket{S_3} = \ket{A}$, we have
\bea
\sqrt{2} A(\bd\to \pi^+\pi^0\pi^-)_{\ket{A}} &=& \left[ 2 T_1^A 
- 2 T_2^A - C_1^A - C_2^A 
- 3 P^A_{uc}  \right] e^{-i\alpha}
\nn\\
&& \hskip-1truein + \left[ 3 P^A_{tc} + P_{EW1}^A  -~P_{EW2}^A 
- P^{C,A}_{EW1} - P^{C,A}_{EW2} \right] ~.
\eea

Now, the final state has isospin $1 \otimes 1 \otimes 1 = 0 \oplus 1
\oplus 1 \oplus 1 \oplus 2 \oplus 2 \oplus 3$.  Given that the
$B$-meson has $I=\frac12$ and the weak Hamiltonian has $\Delta I =
\frac12$ or $\frac32$, there are 9 paths to the final state. We
therefore expect four relations among the 13 decay amplitudes. This is
indeed what is found:
\bea
&& \hskip-3truein \sqrt{2} A(\bd\to \pi^0\pi^0\pi^0)_{\ket{S}} = - \sqrt{3} A(\bd\to \pi^+\pi^0\pi^-)_{\ket{S}} ~, \nn\\
&& \hskip-3truein 2 A(B^+\to \pi^+\pi^0\pi^0)_{\ket{S}} = -A(B^+\to \pi^-\pi^+\pi^+)_{\ket{S}} ~, \nn\\
\frac32 A(\bd\to \pi^+\pi^0\pi^-)_{\ket{M_1}} + \frac{\sqrt{3}}{2}
A(\bd\to \pi^+\pi^0\pi^-)_{\ket{M_3}} &=& \nn\\
&& \hskip-2.5truein A(B^+\to \pi^+\pi^0\pi^0)_{\ket{M_1}} - A(B^+\to \pi^-\pi^+\pi^+)_{\ket{M_1}} ~, \nn\\
\frac32 A(\bd\to \pi^+\pi^0\pi^-)_{\ket{M_2}} + \frac{\sqrt{3}}{2}
A(\bd\to \pi^+\pi^0\pi^-)_{\ket{M_4}} &=& \nn\\
&& \hskip-2.5truein A(B^+\to \pi^+\pi^0\pi^0)_{\ket{M_2}} - A(B^+\to \pi^-\pi^+\pi^+)_{\ket{M_2}} ~.
\label{relations}
\eea
These relations can also be found using the Wigner-Eckart theorem.

In passing, we note that, within the SM, the final state with $I=3$ is
unreachable. This then provides a test of the SM. Applying the method
of Ref.~\cite{Dpipipi} to $B\to\pi\pi\pi$, one can distinguish the
various isospin final states. One can then look for a state with
$I=3$. If one is observed, this will be a smoking-gun signal of new
physics.

\subsection{Dalitz Plots}

Above, we presented the amplitudes for each of the six $S_3$ states of
$B \to \pi\pi\pi$. The obvious question is then whether these states
can be distinguished experimentally. Below we show that this can
indeed be done.

Consider the decay $\bd \to \pi^+\pi^0\pi^-$. The Dalitz-plot events
can be described by $s_+ = \left( p_{\pi^0} + p_{\pi^+} \right)^2$ and
$s_- = \left( p_{\pi^0} + p_{\pi^-} \right)^2$, so that the decay
amplitude, ${\cal M}(s_+,s_-)$, can be extracted. We introduce the
third Mandelstam variable, $s_0 = \left( p_{\pi^+} + p_{\pi^-}
\right)^2$. It is related to $s_+$ and $s_-$ as follows:
\beq
s_+ + s_- + s_0 = m_B^2 + 3m_\pi^2 ~.
\eeq
The totally symmetric SU(3) decay amplitude is then given by
\bea
\ket{S} &\!=\!& \frac{1}{\sqrt{6}} \left[ {\cal M}(s_+,s_-) + {\cal M}(s_-,s_+) +
  {\cal M}(s_+,s_0) \right. \nn\\
&& 
\hskip0.8truecm 
\left. +~{\cal M}(s_0,s_+) + {\cal M}(s_0,s_-) + {\cal
    M}(s_-,s_0) \right] ~.
\eea
Also,
\bea
\ket{M_1} &\!=\!& \frac{1}{\sqrt{12}} \left[ 2{\cal M}(s_+,s_-) +
  2{\cal M}(s_-,s_+) - {\cal M}(s_+,s_0) \right. \nn\\
&& 
\hskip0.8truecm 
\left.  -~{\cal M}(s_0,s_+) - {\cal M}(s_0,s_-) - {\cal
    M}(s_-,s_0) \right] ~.
\eea
The remaining $S_3$ states can be found similarly. The method is similar
for the other $B \to \pi\pi\pi$ decays.

\subsection{Weak-Phase Information}

In the previous subsection we showed how all six $B\to\pi\pi\pi$ $S_3$
states can be experimentally separated.  It may then be possible to
extract clean information about weak phases. (Note: by measuring the
$S_3$ states, one fixes the CP of the final states, which makes the
indirect CP asymmetries well-defined.)

Consider $\ket{S_3} = \ket{A}$. Here there is one decay, which yields
three observables: the branching ratio, the direct CP asymmetry, and
the indirect CP asymmetry of $\bd\to \pi^+\pi^0\pi^-|_{\ket{A}}$. The
amplitude is expressed in terms of two effective diagrams: $A(\bd\to
\pi^+\pi^0\pi^-)_{\ket{A}} = D_1 e^{-i\alpha} + D_2$, which has four
theoretical parameters -- the magnitudes of $D_{1,2}$, the relative
strong phase, and $\alpha$. Since the number of theoretical unknowns
is greater than the number of observables, one cannot obtain
$\alpha$. Things are similar for $\ket{S_3} = \ket{S}$. Due to the
first two relations in Eq.~(\ref{relations}), there are only two independent
decays, yielding 5 observables. However, there are 8 theoretical
parameters, so that, once again, $\alpha$ cannot be extracted.

Things are different for the case of mixed states. Consider the
$M_1$/$M_3$ sector. There are four decays: (1) $B^+\to
\pi^+\pi^0\pi^0|_{\ket{M_1}}$, (2) $B^+\to
\pi^-\pi^+\pi^+|_{\ket{M_1}}$, (3) $\bd\to
\pi^+\pi^0\pi^-|_{\ket{M_1}}$, (4) $\bd\to
\pi^+\pi^0\pi^-|_{\ket{M_3}}$. These yield 10 observables: 4 branching
ratios, 4 direct CP asymmetries, and 2 indirect CP asymmetries (of
$\bd\to \pi^+\pi^0\pi^-|_{\ket{S_3}}$, $S_3 = M_1$, $M_3$). The four
decay amplitudes all have the form $D_{1,i} e^{-i\alpha} + D_{2,i}$,
$i=1$-4. The $D_{1,i}$ are related to one another by the third
relation in Eq.~(\ref{relations}), as are the $D_{2,i}$. The
amplitudes are thus a function of 6 effective diagrams, resulting in
12 theoretical parameters: 6 magnitudes, 5 relative strong phases, and
$\alpha$. Since the number of theoretical unknowns exceeds the number
of observables, $\alpha$ cannot be extracted. However, if one assumes
that the hierarchy of Eq.~(\ref{btodhierarchy}) holds for three-body
decays, all EWP diagrams can be neglected, to a good approximation. In
this case, all the $D_{2,i}$ are proportional to $P^{M_1}_{tc} -
\sqrt{3} P^{M_3}_{tc}$. There are thus only 4 effective diagrams,
which yield 8 theoretical parameters. Now the number of theoretical
unknowns is smaller than the number of observables, so that $\alpha$
can be obtained from a fit to the data. (It is not even necessary to
measure all 10 observables. A difficult-to-obtain quantity, such as
the direct CP asymmetry in $B^+\to \pi^+\pi^0\pi^0|_{\ket{M_1}}$, can
be omitted.) A similar method holds for the $M_2$/$M_4$ sector. The
error on $\alpha$ can be reduced by comparing the two values found.

Now, it must be conceded that the above analysis is quite theoretical
-- it is far from certain that this can be carried out experimentally
[and there is an uncertain theoretical error due to the assumption of
  Eq.~(\ref{btodhierarchy})]. Still, it is interesting to see that, in
principle, clean weak-phase information can be obtained from
$B\to\pi\pi\pi$, or, more generally, from $B \to M_1 M_2 M_3$ decays.

\section{Conclusions}

In this paper, we have expressed the amplitudes for $B \to M_1 M_2
M_3$ decays ($M_i$ is a pseudoscalar meson) in terms of diagrams,
concentrating on the charmless final states $K\pi\pi$, $KK{\bar K}$,
$K{\bar K}\pi$ and $\pi\pi\pi$. The diagrams are similar to those used
in two-body decays: the color-favored and color-suppressed tree
amplitudes $T$ and $C$, the gluonic-penguin amplitudes $P_{tc}$ and
$P_{uc}$, and the color-favored and color-suppressed
electroweak-penguin (EWP) amplitudes $\pew$ and $\pewc$. Here, because
the final state has three particles, there are two types of each
diagram, which we call $T_1$, $T_2$, $C_1$, $C_2$, etc.

We have also demonstrated how to use the Dalitz plots of three-body
decays to separate the decay amplitudes into pieces which are
symmetric or antisymmetric under the exchange of two of the
final-state particles. This is useful for any decay whose final state
contains identical particles under isospin. If the relative angular
momentum of the two particles is even (odd), the isospin state must be
symmetric (antisymmetric). These two possibilities can be
distinguished experimentally.

The main advantage of a diagrammatic analysis is that the approximate
relative sizes of the diagrams can be estimated. For example, there
are annihilation- and exchange-type diagrams which contribute to these
decays. However, these are expected to be negligible, and are not
included in our analysis. Previous studies of three-body decays were
carried out using isospin amplitudes, and gave exact results for the
symmetric or antisymmetric states. On the other hand, the (justified)
neglect of annihilation-type diagrams can modify these results, and
can lead to interesting new effects.

As an example, consider $B \to KK{\bar K}$, which consists of four
decays. For the case where the two $K$'s are in a symmetric isospin
state, the Wigner-Eckart theorem gives a single relation among the
four amplitudes. However, when the amplitudes are written in terms of
the non-negligible diagrams, it is found that this relation actually
consists of two equalities, and this leads to new predictions of the
standard model (SM). Present data allow us to test one of these
equalities, and we find agreement with the SM. In the same vein, $B
\to KK{\bar K}$ decays can be written in terms of five isospin
amplitudes. The diagrammatic analysis shows that, in fact, only four
of these are independent -- two of the isospin amplitudes are
proportional to one another.

Another consequence of the diagrammatic analysis has to do with weak
phases. The CP of a three-particle final state is not fixed, because
the relative angular momenta are unknown (i.e.\ they can be even or
odd). For this reason, in the past it was thought that it is not
possible to cleanly extract weak-phase information from three-body $B$
decays. In this paper, we demonstrate that this is not true. Using the
diagrams, we show that it is possible to cleanly measure the weak
phases in some decays, given that it is experimentally possible to
distinguish different symmetry combinations of the final-state
particles. We explicitly give methods for $K{\bar K}\pi$ and
$\pi\pi\pi$, and note that the the procedure for $K\pi\pi$ is
presented separately. Ways of cleanly extracting the CP phases from
other three-body decays will surely be suggested.

There are thus a number of interesting measurements that can be
carried out with $B \to M_1 M_2 M_3$. LHCb is running at present, and
the super-$B$ factories will run in the future. Hopefully, these
machines will provide interesting data on three-body $B$ decays.

\bigskip
\noindent
{\bf Acknowledgments}:
We thank M. Gronau, J. Rosner, R. Sinha, R. MacKenzie, A. Soffer and
Fran{\c c}oise Provencher for helpful communications, and A. Datta for
collaboration in the beginning stages of this project. This work was
financially supported by NSERC of Canada and FQRNT of Qu\'ebec.




\begin{thebibliography}{99}

\bibitem{hfag}
  E.~Barberio {\it et al.}  [Heavy Flavor Averaging Group],
  arXiv:0808.1297 [hep-ex], and
online update at {\tt http://www.slac.stanford.edu/xorg/hfag}

\bibitem{LNQS} H.~J.~Lipkin, Y.~Nir, H.~R.~Quinn and A.~Snyder,
  Phys.\ Rev.\  D {\bf 44}, 1454 (1991).

\bibitem{GR2003} M.~Gronau and J.~L.~Rosner,
  Phys.\ Lett.\  B {\bf 564}, 90 (2003)
  [arXiv:hep-ph/0304178].

\bibitem{GR2005} M.~Gronau and J.~L.~Rosner,
  Phys.\ Rev.\  D {\bf 72}, 094031 (2005)
  [arXiv:hep-ph/0509155].

\bibitem{Sonietal} H.~Y.~Cheng, C.~K.~Chua and A.~Soni,
  Phys.\ Rev.\  D {\bf 72}, 094003 (2005)
  [arXiv:hep-ph/0506268],
  Phys.\ Rev.\  D {\bf 76}, 094006 (2007)
  [arXiv:0704.1049 [hep-ph]].

\bibitem{GHLR}  M.~Gronau, O.~F.~Hernandez, D.~London and
J.~L.~Rosner, Phys.\ Rev.\ D {\bf 50}, 4529 (1994), Phys.\
Rev.\ D {\bf 52}, 6374 (1995).

\bibitem{K+pi-pi+} A.~Garmash {\it et al.}  [BELLE Collaboration],
  Phys.\ Rev.\  D {\bf 71}, 092003 (2005)
  [arXiv:hep-ex/0412066];
B.~Aubert {\it et al.}  [BABAR Collaboration],
  Phys.\ Rev.\  D {\bf 72}, 072003 (2005)
  [Erratum-ibid.\  D {\bf 74}, 099903 (2006)]
  [arXiv:hep-ex/0507004].

\bibitem{pdg}
  C.~Amsler {\it et al.}  [Particle Data Group],
  Phys.\ Lett.\  B {\bf 667}, 1 (2008).

\bibitem{PV} See, for example, C.~W.~Chiang and D.~London,
  Mod.\ Phys.\ Lett.\  A {\bf 24}, 1983 (2009)
  [arXiv:0904.2235 [hep-ph]].

\bibitem{K+pi-pi+new} B.~Aubert {\it et al.}  [BABAR Collaboration],
  Phys.\ Rev.\  D {\bf 78}, 012004 (2008)
  [arXiv:0803.4451 [hep-ex]];
A.~Garmash {\it et al.}  [Belle Collaboration],
  Phys.\ Rev.\ Lett.\  {\bf 96}, 251803 (2006)
  [arXiv:hep-ex/0512066].

\bibitem{K0pi-pi+} B.~Aubert {\it et al.}  [BABAR Collaboration],
  Phys.\ Rev.\  D {\bf 80}, 112001 (2009)
  [arXiv:0905.3615 [hep-ex]];
A.~Garmash {\it et al.}  [Belle Collaboration],
  Phys.\ Rev.\  D {\bf 75}, 012006 (2007)
  [arXiv:hep-ex/0610081];
J.~Dalseno {\it et al.}  [Belle Collaboration],
  Phys.\ Rev.\  D {\bf 79}, 072004 (2009)
  [arXiv:0811.3665 [hep-ex]].

\bibitem{K+pi0pi-} B.~Aubert {\it et al.}  [BaBar Collaboration],
  Phys.\ Rev.\  D {\bf 78}, 052005 (2008)
  [arXiv:0711.4417 [hep-ex]].

\bibitem{DSS} N.~G.~Deshpande, N.~Sinha and R.~Sinha,
  Phys.\ Rev.\ Lett.\  {\bf 90}, 061802 (2003)
  [arXiv:hep-ph/0207257].

\bibitem{EWPs} M.~Neubert and J.~L.~Rosner,
  Phys.\ Lett.\  B {\bf 441}, 403 (1998)
  [arXiv:hep-ph/9808493],
  Phys.\ Lett.\  B {\bf 441}, 403 (1998)
  [arXiv:hep-ph/9808493];
M.~Gronau, D.~Pirjol and T.~M.~Yan,
  Phys.\ Rev.\  D {\bf 60}, 034021 (1999)
  [Erratum-ibid.\  D {\bf 69}, 119901 (2004)]
  [arXiv:hep-ph/9810482];
M.~Imbeault, A.~L.~Lemerle, V.~Page and D.~London,
  Phys.\ Rev.\ Lett.\  {\bf 92}, 081801 (2004)
  [arXiv:hep-ph/0309061].

\bibitem{Grocomment} M.~Gronau,
  Phys.\ Rev.\ Lett.\  {\bf 91}, 139101 (2003)
  [arXiv:hep-ph/0305144].

\bibitem{Kpipigamma} M.~Imbeault, N.~Rey-Le Lorier and D.~London,
  arXiv:1011.4973 [hep-ph].

\bibitem{KKKBelle} Y.~Nakahama {\it et al.}  [BELLE Collaboration],
  Phys.\ Rev.\  D {\bf 82}, 073011 (2010)
  [arXiv:1007.3848 [hep-ex]].
  
\bibitem{KKKBabar} B.~Aubert {\it et al.}  [BABAR Collaboration],
  Phys.\ Rev.\ Lett.\  {\bf 99}, 161802 (2007)
  [arXiv:0706.3885 [hep-ex]].

\bibitem{Dpipipi} M.~Gaspero, B.~Meadows, K.~Mishra and A.~Soffer,
  Phys.\ Rev.\  D {\bf 78}, 014015 (2008)
  [arXiv:0805.4050 [hep-ph]].

\end{thebibliography}
\end{document}